\documentclass[preprint,12pt,authoryear]{elsarticle}

\usepackage{amssymb}
\usepackage{amsmath}
\usepackage{amsthm}

\usepackage[utf8]{inputenc}
\usepackage{amsfonts}
\usepackage{graphicx} 
\usepackage{tabularx}
\usepackage{float}

\usepackage[labelformat=simple]{subcaption}

\DeclareCaptionLabelFormat{subcaptionlabel}{\normalfont(\textbf{#2}\normalfont)}
\usepackage{setspace}
\usepackage{booktabs}
\usepackage{multirow}
\usepackage{caption}  
\usepackage{placeins}  
\usepackage{adjustbox}
\usepackage{fancyvrb}
\usepackage{natbib}
\usepackage{xcolor}
\usepackage{hyperref}
\usepackage[english]{babel}
\usepackage{ragged2e}
\usepackage{blindtext}
\usepackage{wrapfig}
\usepackage{fancyhdr}
\usepackage{geometry}
\usepackage[flushleft]{threeparttable}
\hypersetup{
  colorlinks=true,
  linkcolor=blue,  
  urlcolor=red,
  pdftitle={}
}
\usepackage{soul}

\def\Si{\mathsf S}
\def\Ei{\mathsf E}
\def\Ii{\mathsf I}
\def\Ri{\mathsf R}
\def\Ni{\mathsf N}
\def\R{\mathcal{R}}
\def\Ro{\R_{0}}
\def\dd{\mathrm{d}}

\usepackage{titlesec}
\titleformat{\paragraph}
{\normalfont\normalsize\it}{\theparagraph}{1em}{}
\titlespacing*{\paragraph}
{0pt}{3.25ex plus 1ex minus .2ex}{0pt}
\usepackage{indentfirst}

\journal{Epidemics}

\begin{document}

\begin{frontmatter}



\title{On the simultaneous inference of susceptibility distributions and intervention effects from epidemic curves}

\author[1,2]{Ibrahim Mohammed}
\author[1,3]{Chris Robertson}
\author[1,4]{M. Gabriela M. Gomes\corref{*}}

\cortext[*]{Corresponding author: gabriela.gomes@strath.ac.uk}

\affiliation[1]{organization={Department of Mathematics and Statistics, University of Strathclyde},
            city={Glasgow},
            postcode={G1 1XH}, 
            country={UK}}
\affiliation[2]{organization={Department of Mathematical Sciences, Abubakar Tafawa Balewa University},
            city={Bauchi},
            country={Nigeria}}
\affiliation[3]{organization={Public Health Scotland},
            city={Glasgow},
            postcode={G2 6QE}, 
            country={UK}}
\affiliation[4]{organization={Centre for Mathematics and Applications (NOVA MATH), NOVA School of Science and Technology},
            city={Caparica},
            postcode={2829-516}, 
            country={Portugal}}

\begin{abstract}
Susceptible-Exposed-Infectious-Recovered (SEIR) models with inter-individual variation in susceptibility or exposure to infection were proposed early in the COVID-19 pandemic as a potential element of the mathematical/statistical toolset available to policy development. In comparison with other models employed at the time, those designed to fully estimate the effects of such variation tended to predict small epidemic waves and hence require less containment to achieve the same outcomes. However, these models never made it to mainstream COVID-19 policy making due to lack of prior validation of their inference capabilities. Here we report the results of the first systematic investigation of this matter. We simulate datasets using the model with strategically chosen parameter values, and then conduct maximum likelihood estimation to assess how well we can retrieve the assumed parameter values. We identify some identifiability issues which can be overcome by creatively fitting multiple epidemics with shared parameters. 
\end{abstract}




\end{frontmatter}

\section{Introduction}

Susceptible-Exposed-Infectious-Recovered (SEIR) models are used extensively to study epidemics and guide public health policies. These models range in detail, from simple systems of ordinary differential equations (ODEs), to higher-dimensional implementations that include disease progression, age structure, and other forms of heterogeneity \citep{Diekmann2013}. Metapopulation \citep{Keeling2002} or agent-based models \citep{Kerr2021} are most commonly used when explicit descriptions of spatial connectivity are desired, although they also enable the representation of other heterogeneities such as person-to-person variability as stylised in \citep{Ayabina2025}.

During the COVID-19 pandemic, several authors highlighted the significant role of individual variation in susceptibility and exposure to infection in flattening epidemic curves \citep{Britton2020, Neipel2020, Rose2021, Tkachenko2021, Montalban2022, Gomes2022}. However, despite being understood for decades \citep{Mckendrick1940, Gart1968, Gart1972, Ball1985, Katriel2012}, these ideas were treated with caution among scientific advisors to governments due to supposed parameter identifiability issues \citep{Wood2025,GMR2025}. Here we report the results of a thorough investigation of parameter identifiability in SEIR models.

We use SEIR models with individual variation in susceptibility to infection proposed in \citep{Montalban2022, Gomes2022} during the pandemic. These models include non-pharmaceutical interventions (NPIs) as required, which also tend to flatten epidemic curves and be potentially confounded with heterogeneity in susceptibility. To test our ability to attribute flattning of epidemic curves correctly to heterogeneity or NPIs we generate synthetic epidemic datasets by simulating stochastic versions of those models. Then we perform sets of statistical inferences by fitting homogeneous and heterogeneous versions of the models to the simulated data. Model parameters are estimated by maximum likelihood.

In our study setup, we find that models that allow susceptibility to vary among individuals are able to infer that the coefficient of variation (CV) is negligible when fitting a dataset generated by a homogeneous model with NPIs. All other parameters are also accurate. By applying the Akaike information criterion (AIC), we obtain similar scores for homogeneous and heterogeneous models when the datasets are generated under the homogeneity assumption. By contrast, when the datasets are generated with inter-individual variation in susceptibility, homogeneous models overestimate the impact of NPIs as misattribution of the effects of heterogeneity. AIC scores are noticeably lower for homogeneous models in these scenarios.

In addition to assessing the accuracy of the estimated parameters, we noticed some strong correlations between parameters. This was irrespective of whether models accounted for individual variation or not. To generate further insight, we varied the initial number of infectious individuals (seed) as to gain control over the stage of the epidemic at the point of NPI introduction, we developed a scheme to fit two epidemic curves with different seeds (akin metapopulation modelling \citep{Ayabina2025}), and conducted profile likelihood analyses. We conclude that two-epidemic fittings (as performed in \citep{Gomes2022}) are an effective strategy for reducing parameter correlations, and hence overcoming identifiability issues, overall.

\section{Mathematical models}

\subsection{SEIR models with inter-individual variation in susceptibility}\label{seir_section}

We adopt an SEIR model previously analysed by \cite{Montalban2022} and applied to the COVID-19 pandemic by \cite{Gomes2022}. Inter-individual variation in susceptibility to infection is incorporated as a multiplicative factor, $x$, of the rate of infection. The model is written in terms of differential equations as
\begin{equation}\label{Hetro_system}
\begin{split}
 \frac{\dd S(x)}{\dd t}&= - c(t)\ \beta\ (\rho\ \Ei+\Ii)\ \frac{x\ S(x)}{\Ni}, \\
 \frac{\dd\Ei}{\dd t}&=c(t)\ \beta\ (\rho\ \Ei+\Ii)\ \int  \frac{x\ S(x)}{\Ni}\ {\dd x} - \delta\ \Ei, \\
 \frac{\dd\Ii}{\dd t}&= \delta\ \Ei-\gamma\ \Ii,
\end{split}
\end{equation}
where $S(x)$ represents the density of susceptible individuals as a function of the susceptibility factor $x$, $\Ei+\Ii$ is the number of individuals who have been exposed and are infected (accounting for an early stage of lower infectiousness, $\Ei$). The recovered number $\Ri$ is derived from the conservation of total population size, $\Si+\Ei+\Ii+\Ri=\Ni$, where $\Si=\int S(x)\ \dd x$. The main parameters are the average effective contact rate $\beta$, the rate of progression from $\Ei$ to $\Ii$ (assumed $\delta=1/5.5$ per day \citep{McAloon2020,Lauer2020}), the rate of removal from $\Ii$ (assumed $\gamma=1/4$ per day \citep{Nishiura2020,Li2020}) and the reduced infectiousness while in $\Ei$ (assumed $\rho = 0.5$). An additional time-dependent parameter $c(t)$ is included, which has a default value of $1$ and will be used to study the effect of interventions. Initial conditions for variables $S(x,t)$, $\Ei(t)$, $\Ii(t)$ are defined as to satisfy $S(x,0)=(1-3.5\ \epsilon)\ q(x)\ \Ni$, $\Ei(t)=2.5\ \epsilon\ \Ni$ and $\Ii(t)=\epsilon\ \Ni$ (see Supplementary material Section S1 for justification), where $q(x)$ is a probability density function with mean 1 and coefficient of variation
\begin{equation}\label{coeff_var}
    \nu =\sqrt{\int (x-1)^2 q(x)dx}.
\end{equation}

The basic reproduction number is given by
\begin{equation}\label{r0gomes22}
\Ro= \beta\left(\frac{\rho}{\delta}+\frac{1}{\gamma}\right).
\end{equation}

Following \cite{Novozhilov2008}, \cite{Montalban2022} showed that when $q(x)$ is a gamma distribution, system \eqref{Hetro_system} can be reduced exactly to
\begin{equation}\label{Reduced_system}
\begin{split}
 \frac{\dd \Si}{\dd t}&= -c(t)\ \beta\ (\rho\ \Ei+\Ii)\ \left(\frac{\Si}{\Ni} \right)^{1+\nu^2}, \\
 \frac{\dd \Ei}{\dd t}&= c(t)\ \beta\ (\rho\ \Ei+\Ii)\ \left(\frac{\Si}{\Ni} \right)^{1+\nu^2} - \delta\ \Ei, \\
\frac{\dd \Ii}{\dd t}&= \delta\ \Ei-\gamma\ \Ii,
\end{split}
\end{equation}
The gamma distribution was described by \cite{Rose2021} as a ``natural choice to account for variations in susceptibility'' as it is the limiting distribution to which other initial distributions of susceptibility converge. This choice will be kept throughout this paper, although the treatment of the explicit system \eqref{Hetro_system} can be replicated for any distribution while the use of the reduced version \eqref{Reduced_system} is specific to gamma distributed susceptibility. In either formulation, the classical SEIR model is retrieved by setting $\nu=0$.

\begin{figure}[!htb]
    \centering
    \includegraphics[width=0.8\linewidth]{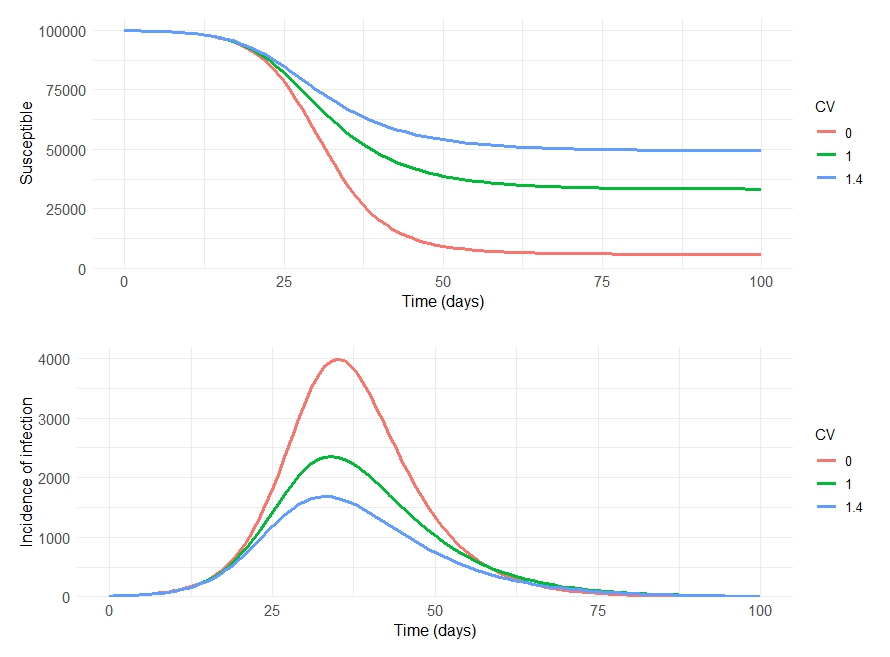}
    \caption{Dynamics of infection under the heterogeneous susceptibility SEIR model for different values of \(\nu\) (bottom). The homogeneous SEIR model corresponds to \( \nu=0\). The top figure shows how the dynamics of the susceptible pool size changes with \( \nu\). }
    \label{fig:coeffovarcurves}
\end{figure}

Previous studies, including those in \citep{Novozhilov2008,Neipel2020,Gomes2022,Montalban2022,Rose2021}, have demonstrated that heterogeneity in susceptibility reduces epidemic sizes. This occurs because highly susceptible individuals are infected first and hence selected out of the susceptible pool, a dynamic not captured by the classical homogeneous model. This effect is illustrated in Figure~\ref{fig:coeffovarcurves} (top). For example, around $50\%$ of the population is expected to be infected over the course of an outbreak when CV is $\nu=\sqrt{2}\approx 1.414$. By comparison, when a homogeneous model is assumed ($\nu=0$) the expected infected percentage rises to around $90\%$. In the bottom panel, corresponding trajectories for incidence of infection (represented as $\delta\ \Ei$) are shown, highlighting the higher peak in the absence of individual variation.

\FloatBarrier 

\subsection{Non-pharmaceutical interventions  (NPIs)}\label{npi_section}

\begin{figure}[!htb]
    \centering
    \includegraphics[trim={0 1cm 0 1cm},clip,width=0.8\textwidth]{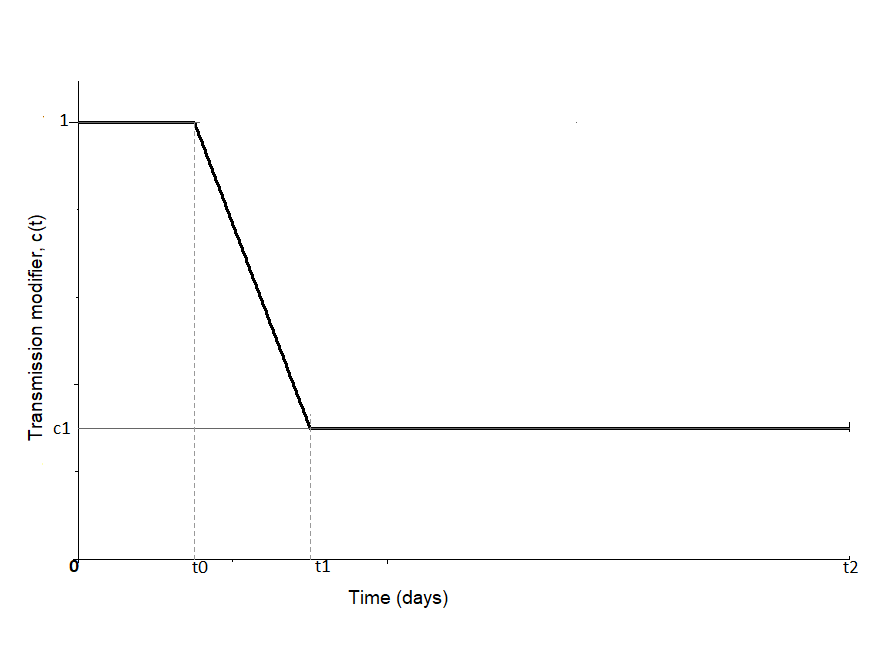}
    \caption{Schematic illustration of the factor 
\( c(t)\), representing the combined effects of adaptive behavioural changes and NPIs on transmission.}
   \label{fig:NPIprofile}
\end{figure}

Outbreak awareness triggers, to some extent, a reduction in contract rates. This may be due to voluntary precautions or restrictions imposed by governments. To incorporate the effects of these co-called non-pharmaceutical interventions (NPIs) we use a time dependent factor \( c(t)\) such as
\begin{equation}
c(t) =
\begin{cases} \label{npi}
  1, & \text{if } 0 < t \leq t_0 \\
  \displaystyle 1 - (1-c_1)\frac{t-t_0}{t_1-t_0}, & \text{if } t_0 < t \leq t_1 \\
  c_1, &  \text{if  } t_1 < t 
\end{cases}
\end{equation}
where $0\leq c_1\leq 1$. This is profiled in Figure~\ref{fig:NPIprofile}, with \( t_0\) representing the time when contact rates begin to decrease and \(t_1\) marking the beginning of maximal containment (such as lockdown).

\begin{figure}[!htb]
    \centering
    \includegraphics[width=0.8\linewidth]{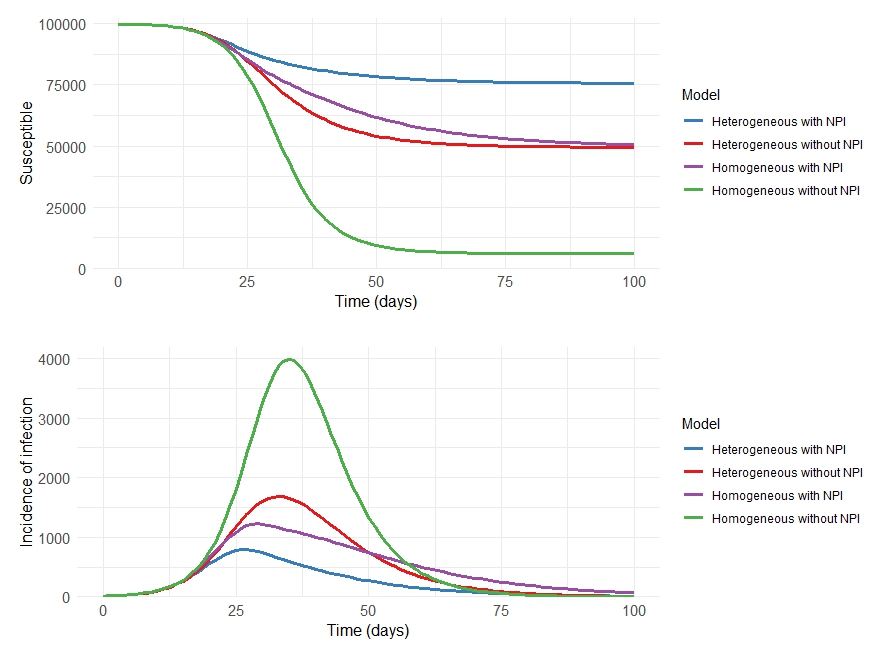}
    \caption{ The combined effects of heterogeneity in susceptibility and NPIs on epidemic trajectories for different values of \(\nu\) and constant NPI \(c_1=0.4\) (bottom). The homogeneous SEIR model corresponds to \( \nu=0\). The top figure shows how the susceptible pool changes with both \(\nu\) and  \( c_1\).}
    \label{fig:Combined_curves_withNPIs}
\end{figure}

When $c_1<1$, the dynamics of the susceptible numbers and the incidence of infection for a homogeneous model are affected as shown in Figure~\ref{fig:Combined_curves_withNPIs}. It is clear that both
heterogeneity in susceptibility and NPIs flatten epidemic curves although for the specific NPI profiles adopted here the flattened curves have distinctive shapes. 

A description of all the model parameters is provided in Table~\ref{tab1}.

\begin{table}[!htb]
\centering
\caption{Description of model parameters. \label{tab1}}
\newcolumntype{C}{>{\centering\arraybackslash}X}
\begin{tabularx}{\textwidth}{>{\centering\arraybackslash}m{3cm}C>{\centering\arraybackslash}m{3cm}}
\toprule
\textbf{Parameter} & \textbf{Description} & \textbf{Value} \\
\midrule
$\Ni$ & Total population & $100,000$ \\
$\delta$ & Rate of progression from $\Ei$ to $\Ii$ & $1/5.5$ per day \\
$\gamma$ & Rate of removal from $\Ii$ & $1/4$ per day \\
$\rho$ & Infectiousness in $\Ei$ (relative to $\Ii$) & $0.5$ \\
$\Ro$ & Basic reproduction number & Variable \\
$ \nu $ & Coefficient of variation in susceptibility & Variable\\
$c_1$ & Maximal reduction in transmission by NPIs &  Variable\\
$t_0$ & Time to beginning of behavioural change & 15 days \\
$t_1$ & Time to beginning of lockdown & $20$ days \\
\bottomrule
\end{tabularx}
\end{table}

\section{Simulations} \label{datasets}

We aim to identify the effects of individual variation and NPIs on epidemic trajectories based on a given epidemic dataset affected by both. We design a systematic study with four different incidence time series generated as follows:
 \begin{enumerate}  
    \item Homogeneous model ($\nu=0$) without NPIs ($c_1=1$). 
    \item Heterogeneous model ($\nu>0$) without NPIs ($c_1=1$).
     \item Homogeneous model ($\nu=0$) with NPIs ($c_1<1$).
    \item  Heterogeneous model ($\nu>0$) with NPIs ($c_1<1$).
\end{enumerate}
The datasets are generated by running model \eqref{Reduced_system}, with NPIs as in \eqref{npi}. We define incidence of infection as $\delta\ \Ei$ and adopt a Poisson error structure to obtain the time series
\[ y(t) \sim \mbox{Poisson} ( \hat{\iota}(t) ) ~~~~  \mbox{for each day}~~ t \in \mathbb{N}. \]
 
For each combination of parameters, 200 datasets are simulated and we fit both the homogeneous model (to estimate $\Ro$, $t_0$ and $c_1$) and the model with heterogeneity represented by parameter $\nu$ (to estimate $\Ro$, $\nu$, $t_0$ and $c_1$), while keeping other parameters fixed as in Table~\ref{tab1}. 

\section{Maximum likelihood estimation} \label{Mfit_and_par_est}

With the heterogeneous susceptibility models, we estimate two sets of parameters from the datasets described in the previous section:
\begin{equation} \label{tita1}
    \boldsymbol{\theta_a} :=( \Ro, \nu)
\end{equation}
\begin{equation} \label{tita2}
    \boldsymbol{\theta_b} :=( \Ro, \nu, t_0, c_1).
\end{equation}

The first case corresponds to estimating \( \boldsymbol{\theta_a}\) by fitting the models with no NPIs to the first two datasets in Section~\ref{datasets}. This is to establish how well heterogeneity can be estimated in the absence of the potentially confounding effects of interventions, and to test how reliable the heterogeneous susceptibility model is at estimating ``no variation'' when the dataset has been generated by the homogeneous model.

The second case of interest is estimating \( \boldsymbol{\theta_b}\) from the last two datasets in Section~\ref{datasets}, but we shall also estimate \(\boldsymbol{\theta_b}\) from the other datasets to test how well models recognise when there is ``no intervention''.

Let \( x_1, x_2,...,x_n\) be the model simulation of the observations  for the time series epidemic data \( y_1, y_2,...,y_n\). Let \( f(y_1,y_2,...y_n|x_1,x_2,...,x_n, \boldsymbol{\theta} )\) be  the joint density function which defines a probability distribution for each value of a parameter vector \( \boldsymbol{\theta}:= \boldsymbol{\theta_a} \cup \{ \rho, \delta, \gamma, t_1\}\) or \( \boldsymbol{\theta}:= \boldsymbol{\theta_b} \cup \{ \rho, \delta, \gamma, t_1\}\).

The likelihood function is the density function evaluated at the data which is defined as
\[ \mathcal{L} (\boldsymbol{\theta}):= \prod_{i=1}^n f(y_i | x_i, \boldsymbol{\theta}). \]
It is usually convenient to work with the log-likelihood function given by
\begin{equation} \label{MLE}
   l(\boldsymbol{\theta}):= \log  \mathcal{L} (\boldsymbol{\theta})= \sum_{i=1}^n f(y_i | x_i, \boldsymbol{\theta}).
\end{equation}
Parameter estimation was performed in Rstudio by maximizing the log-likelihood  function \eqref{MLE} of observing the simulated data \( \{y_1,y_2,...,y_n\}\) given the model, and its parameters.

\section{Baseline analysis} \label{baseline}

This section is entirely devoted to presenting and discussing the results of a detailed parameter estimation study for a representative choice of parameters. We group the datasets described in Section~\ref{datasets} into two broad cases:
\begin{enumerate}
    \item[I]: Epidemic datasets simulated from the homogeneous model (\eqref{Reduced_system} with $\nu=0$) and \(\Ro=3\), which are further classified as datasets without intervention ((a) \( c_1=0\)) and with intervention ((b) \( c_1=0.3\)).
    \item[II]: Epidemic datasets simulated from heterogeneous susceptibility model (\eqref{Reduced_system} with $\nu=1.414$) and \(\Ro=3\), which are also classified as datasets without intervention ((a) \( c_1=0\)) and with intervention ((b) \( c_1=0.3\)).
\end{enumerate}
As outlined in Section~\ref{Mfit_and_par_est}, we then estimate parameter combinations \( \boldsymbol{\theta_a} :=( \Ro, \nu)\) or \( \boldsymbol{\theta_b} :=( \Ro, \nu, t_0, c_1)\) from the simulated datasets. Table~\ref{tab:summary_results} shows the resulting summary statistics which are discussed in the remaining of this section. We refer to this as the ``baseline analysis'' as it provides a basis for later sections which either deepen the study or extended it by modifying conditions stipulated here.

\begin{table}[!htb]
\centering
\caption{Summary results for parameter estimates and 95\% confidence intervals. \label{tab:summary_results}}
\resizebox{\textwidth}{!}{%
\begin{tabular}{|c|c|c|c|c|c|}
\hline
\multirow{2}{*}{Case} & \multicolumn{5}{c|}{Parameter estimates and 95\% CI for both Heterogeneous and \textbf{Homogeneous} (\textbf{bold face}) models} \\
\cline{2-6}
 & $\Ro$ & $\nu$ & $t_0$ & $c_1$ & AIC \\
\hline
I(a)(i) & 3.00 (2.99, 3.01) & 0.029 (0.00, 0.33) & NA & NA & 817.23 \\
 & \textbf{3.00 (3.00, 3.01)} & NA & NA& NA & \textbf{815.58} \\
\hline
I(a)(ii) & 3.00 (2.99, 3.03) & 0.0 (0.00, 4.9e219) &0.1 (0.00, 4.6e12)& 0.997 (0.98, 1.00) & 805.71 \\
 & \textbf{3.00 (2.99, 3.00)} & NA & \textbf{0.00 (0.00, 1.05e79)} & \textbf{0.999,(0.784, 1.00)} & \textbf{808.03} \\
\hline 
I(b)& 3.00 (2.97, 3.04) & 0.201 (0.04, 0.912) & 14.9 (14.30, 15.60) & 0.302 (0.296 , 0.308) & 794.73 \\
  & \textbf{3.00 (2.97, 3.04)} & NA & \textbf{15.0(14.4, 15.60)} & \textbf{0.30 (0.296, 0.304)} & \textbf{791.09} \\
\hline
II(a)(i) & 3.00 (2.99, 3.01) & 1.415 (1.40, 1.43) & NA & NA & 814.67 \\
 & \textbf{2.93 (2.92, 2.94)}  & \textbf{NA} & \textbf{NA} & \textbf{NA} & \textbf{30415.99} \\
\hline
II(a)(ii) & 3.00 (2.98, 3.02) & 1.411 (1.394, 1.426) & 0.3 (0, 1.39e21) & 0.995 (0.82, 1.00) & 816.37 \\
& \textbf{2.74 (2.73, 3.00)} & NA & \textbf{  28.5,(28.3, 28.7)} & \textbf{0.352 (0.348, 0.355)} & \textbf{2132.95} \\
\hline 
II(b) & 3.00 (2.95, 3.05) & 1.406(1.157, 1.768) & 15.00 (14.4, 15.70) & 0.299 (0.274, 0.331) & 713.27 \\
 & \textbf{2.94 (2.89, 3.00)} & NA & \textbf{14.90 (14.3, 15.5)} & \textbf{0.24 (0.236, 0.243)} & \textbf{727.54} \\
\hline
\end{tabular}%
}
\end{table}
\subsection{Case I(a): Homogeneous datasets without NPIs} \label{caseIa}

Illustrative fittings for this case are provided in Supplementary Figure S2. There are essentially two scenarios.

First, to test whether the heterogeneous susceptibility model correctly infers that the data had been generated by a homogeneous model, we conduct an analysis where fittings are conducted with models that do not include NPIs (so, only parameters \( \boldsymbol{\theta_a}\) are estimated). This is to test for potential confounding issues between $\Ro$ and $\nu$. Table~\ref{tab:summary_results}, Case I(a)(i), shows the summary statistics for the model-based parameter estimates. Both homogeneous and heterogeneous models give unbiased estimates of the basic reproduction number \( {\Ro}\) ($3.00$, with high precision), with the heterogeneous susceptibility model estimating a coefficient of variation \( {\nu}\) as $0.029$ with $95\%\ \textrm{CI}$ $(0.00,0.33)$. The AIC scores from the table are $815.58$ for homogeneous and $817.23$ for heterogeneous, which are not significantly different. 

Second, to exclude potential confounding issues between $\Ro$ and $c_1$, we use the same simulated datasets as in the first case, but fit them with a model that allows for NPI effects. Table~\ref{tab:summary_results}, Case I(a)(ii), shows the summary statistics for the model-based estimates of parameters \( \boldsymbol{\theta_b}\). Again, both homogeneous and heterogeneous susceptibility models give unbiased estimates of the basic reproduction number \({\Ro}\) ($3.00$, with high precision), with heterogeneous susceptibility also estimating a coefficient of variation \( {\nu}\) very precisely as $0.0$. Furthermore, the NPI parameter, specified as \( c_1=1\) in the simulation of the data, is estimated by both models to be very close to 1, while the estimate for the time point when transmission begins to decrease \( {t_0}\) is estimated as a redundant parameter by both models with extreme confidence intervals. The AIC score is $805.71$ and $808.03$ for homogeneous and heterogeneous models, respectively, so again similar. 



\subsection{Homogeneous datasets with NPIs: Case I(b)} \label{caseIb}

This case is particularly informative given that both heterogeneity and NPIs have the effect of flattening epidemic curves and it is important that any such effects are attributed to the true cause. Illustrative fittings are shown in Supplementary Figure S3.

\begin{figure}[!htb]
    \centering
    \includegraphics[width=0.9\linewidth]{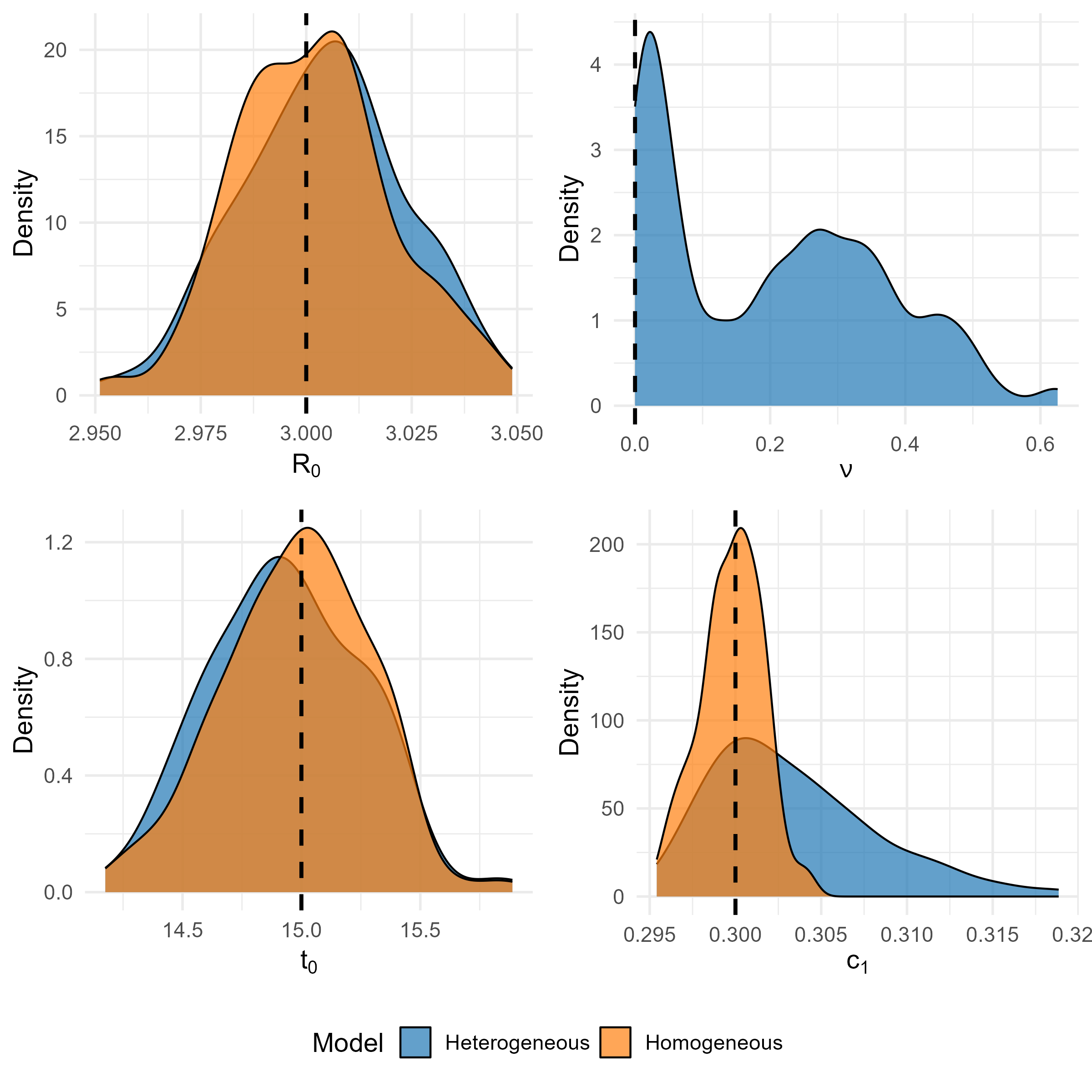}
    \captionsetup{width=0.9\linewidth, justification=justified}
    \caption{Distributions of the estimated parameters for Case I(b): Homogeneous data with impact of NPIs ($\Ro=3$, $c_1=0.3$, $t_0=15$ days). The blue fill represents estimates from the heterogeneous susceptibility model, while the orange fill represents those from the homogeneous model. Vertical dashed lines indicate the true specified parameter values. Both models accurately recover $\Ro$, $t_0$, and the value of the NPI parameter $c_1$, and the heterogeneous model estimates a coefficient of variation $\nu$ near 0.}
    \label{fig:Par_Dist_Case_I__b}
\end{figure}

When we fit the simulated datasets with homogeneous and heterogeneous models with NPIs to estimate parameters \( \boldsymbol{\theta_b}\), both models give similar estimates for the transmission parameters (\(\Ro\), \(c_1\), and \(t_0\)), which are unbiased as can be seen from Table~\ref{tab:summary_results}, Case I(b), and Figure~\ref{fig:Par_Dist_Case_I__b}. The estimates based on the heterogeneous susceptibility (vs. homogeneous) model are \({\Ro}=3.00\) \((3.00)\), \({c_1}=0.302\) \((0.30)\) and \({t_0}=14.90\) \((15.0)\). The true value for CV is $\nu=0$ but the heterogeneous model estimates a biased value \(0.204\) with $95\%\ \textrm{CI}$ $(0.04,0.912)$, as clearly visible in the CV distribution in Figure~\ref{fig:Par_Dist_Case_I__b}. This bias reflects the confounding between NPIs and heterogeneity effects, which we address using the concurrent epidemic approach discussed in later sections. The AIC scores for the two models are similar, with $794.73$ for heterogeneous susceptibility and $791.09$ for the homogeneous model. 

\begin{figure}[!htb]
    \centering
    \includegraphics[width=1\linewidth]{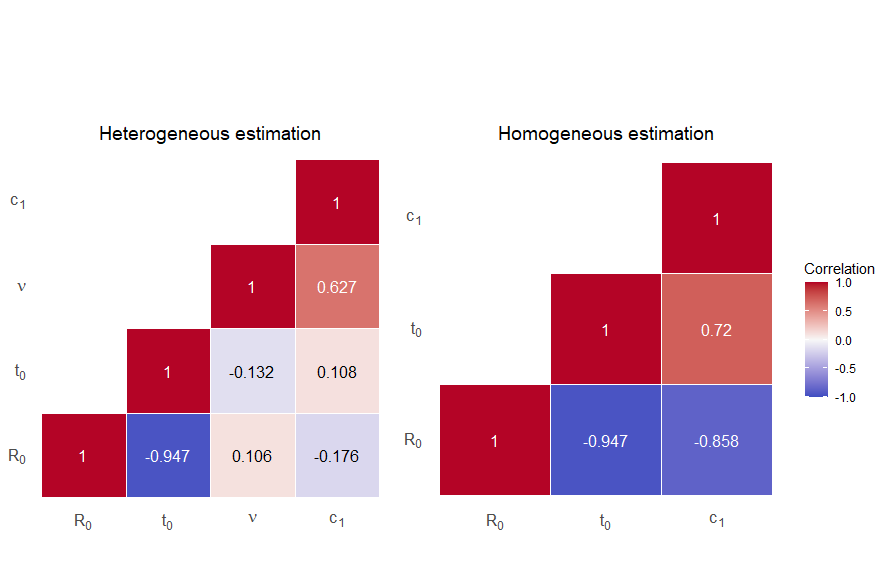}
    \captionsetup{width=0.9\linewidth, justification=justified}
    \caption{Median parameter correlation heatmaps for Case I(b): homogeneous data with impact of NPIs. Both heterogeneous (left) and homogeneous (right) models show very strong negative correlations between $\Ro$ and $t_0$ ($-0.947$). }
    \label{fig:corr_Case_I__b}
\end{figure}

Moreover, from the correlation plots in Figure~\ref{fig:corr_Case_I__b} we identify various correlations between parameters of both models suggesting potential issues with identifiability which we will address in later sections.

Understanding the uncertainty in parameter estimates is also crucial for predicting the potential course of outbreaks. Using the estimated parameters for both homogeneous and heterogeneous models, we constructed prediction intervals and examined the models' forecasting capabilities. We used a multivariate normal distribution for generating parameter sets that preserve the correlations between parameter estimates. This approach allows us to assess not only how well models fit observed data, but crucially, how reliably they predict future epidemic trajectories.

\begin{figure}[ht]
    \centering
    \includegraphics[trim={0 1.5cm 0 0cm},clip,width=\textwidth]{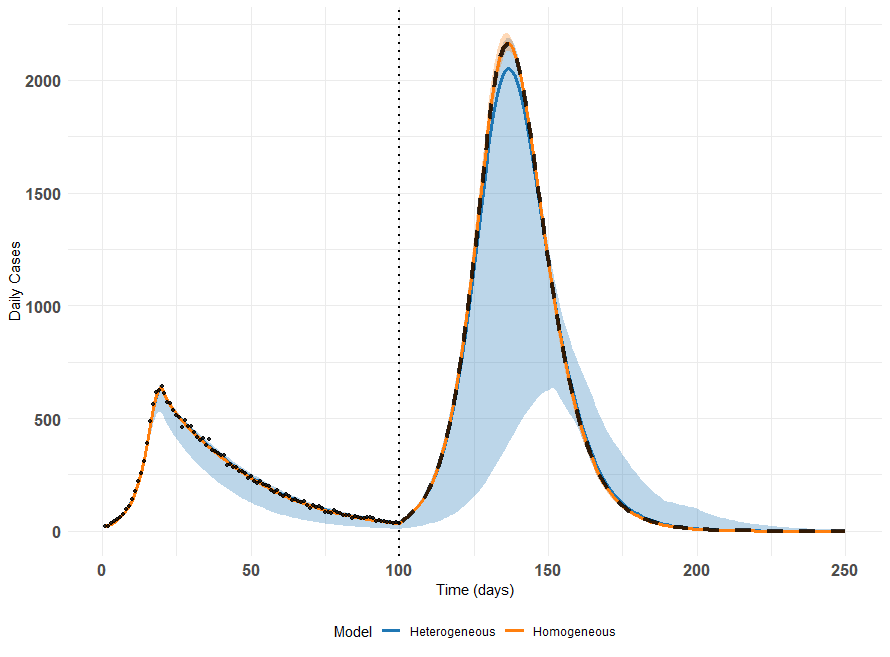}
    \caption{Prediction trajectories for Case I(b): Homogeneous data with NPIs. Models were fitted to the first 100 days of data (black dots) and then used to forecast the next 150 days. Heterogeneous model fit (blue); homogeneous model fit (orange). The dashed black line represents the ``true'' trajectory generated with the assumed parameter values.}
    \label{fig:case_I_b_prediction}
\end{figure}

Figure~\ref{fig:case_I_b_prediction} shows the prediction results for this case. In addition to accurately capturing the trajectory of the epidemic in the fitting period (days 1-100), both models are also good at forecasting beyond that (days 100-250). This is despite the heterogeneous model having wrongly estimated a small but non-zero coefficient of variation. The wide confidence band for the heterogeneous model is mainly due to the uncertainty in the estimation of CV, which has a $95\%\ \textrm{CI}$ of $(0,04, 0.912)$.

\subsection{Case II(a): Heterogeneous datasets without NPIs} \label{caseIIa}

Illustrative fittings for this case are provided in Supplementary Figure S4. There are two scenarios.

First, recalling that heterogeneity in susceptibility flattens epidemic curves, we might expect that, once the dataset has been generated using a model that incorporated some degree of inter-individual variation, it should be very difficult for a homogeneous model to fit the data unless the lack of heterogeneity could be somehow compensated by a change in $\Ro$. But in Section~\ref{caseIa} we did not detect any confounding between $\Ro$ and $\nu$, suggesting that the homogeneous model may fail to provide adequate fits to heterogeneous data when NPIs are not allowed to be part of the inference. This is confirmed in Table~\ref{tab:summary_results}, Case II(a)(i), where the summary statistics for the estimates of model parameters \( \boldsymbol{\theta_a}\) are given. The heterogeneous model recaptures the specified  \( \Ro\) and \( \nu\) as $3.00$ with $95\%\ \textrm{CI}$ $(2.99,3.01)$ and $1.415$ with $95\%\ \textrm{CI}$ $(1.40,1.43)$, respectively, while the homogeneous model gives a biased estimate of the basic reproduction number \( \Ro\) as $2.93$ with $95\%\ \textrm{CI}$ $(2.92,2.94)$. The AIC scores from the table are $814.67$ for heterogeneous susceptibility and $30415.99$ for the homogeneous model, which are far from being close, and indicate that the homogeneous model does not fit the data which is also seen in Figure S4.

Second, we fit the same simulated datasets with a model that allows for NPI effects. We might expect NPIs to provide a means to improve the fits attempted above at the expense of biasing estimates. Table~\ref{tab:summary_results}, Case II(a)(ii), shows the summary statistics for the estimates of model parameters \( \boldsymbol{\theta_b}\). The parameter estimates confirm that the heterogeneous susceptibility model effectively estimates \( \Ro\)  and \( {\nu}\) as $3.00$ with $95\%\ \textrm{CI}$ $(2.98,3.02)$ and $1.411$ with $95\%\ \textrm{CI}$ $(1.394,1.426)$, as well the impact of NPIs \({c_1}\) as \(0.995\) ($95\%\ \textrm{CI}$ being $(0.82,1.00)$ ) - an indication of no effects of NPIs. As for the homogeneous model, in addition to biased estimates of \(\Ro\) as $2.74$ with $95\%\ \textrm{CI}$ $(2.73,3.0)$, it also estimates that there was a strict NPI with \( {c_1}\) as \(0.352\) and $95\%\ \textrm{CI}$ $(0.348,0.355)$ and a non-existent parameter \( t_0\) estimated with high precision, thereby misattributing effect of heterogeneity to an intervention. The AIC scores in this case are $818.07$ for heterogeneous susceptibility and $2132.95$ for the homogeneous model, again confirming the substantially better fit of the heterogeneous model.


\subsection{Case II(b): Heterogeneous datasets with NPIs} \label{caseIIb}

Here we finally generate datasets using the full heterogeneous model \eqref{Reduced_system} and NPIs \eqref{npi}. We aim to estimate both the effect of heterogeneity and the impact of the NPIs from this data. Illustrative fittings are shown in Supplementary Figure S5.

\begin{figure}[!htb]
    \centering
    \includegraphics[width=0.9\linewidth]{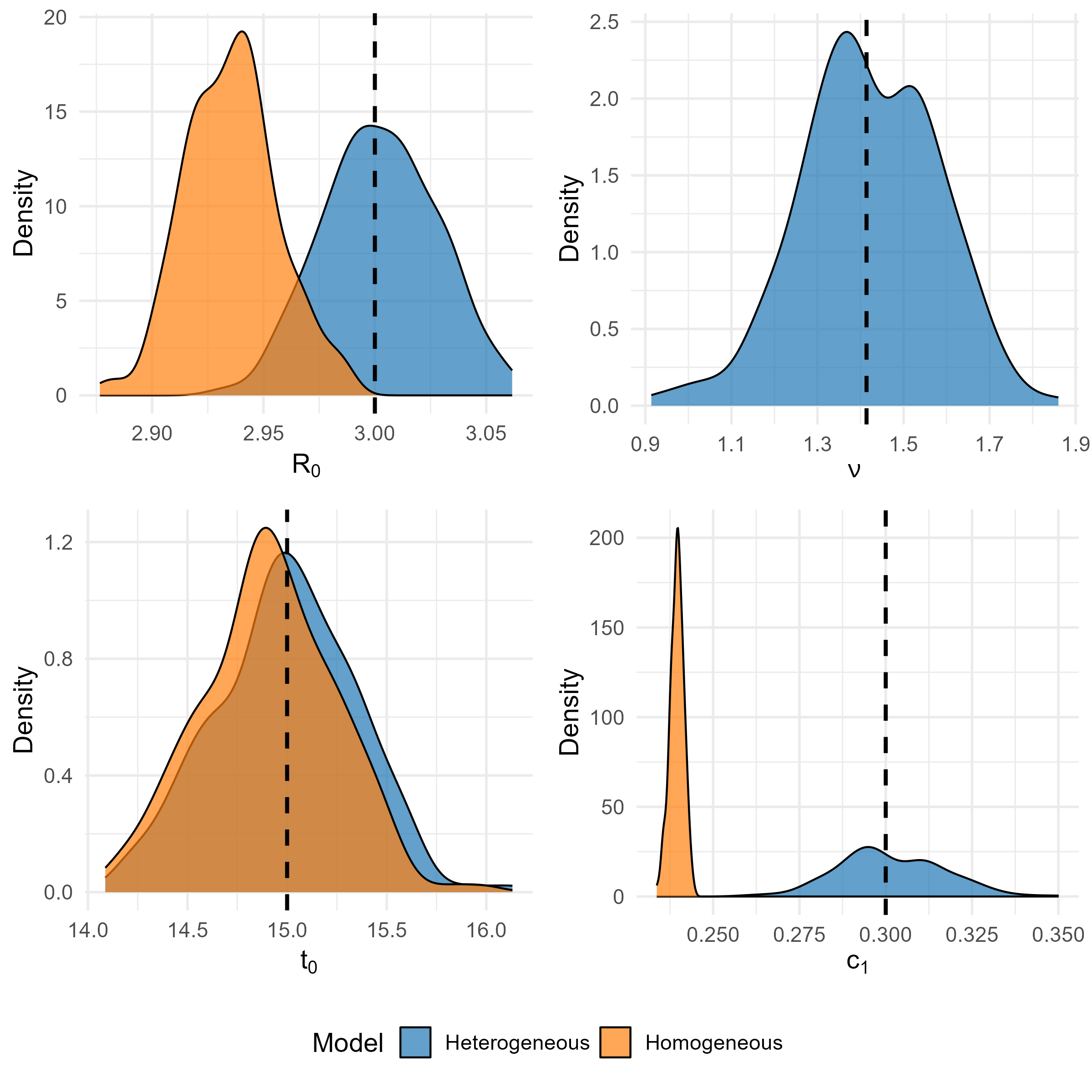}
    \captionsetup{width=0.9\linewidth, justification=justified}
    \caption{{Distributions of the estimated parameters for Case II(b): Heterogeneous data with impact of NPIs ($\Ro=3$, $\nu=1.414$, $c_1=0.3$, $t_0=15$ days).} The blue fill represents estimates from the heterogeneous susceptibility model, while the orange fill represents those from the homogeneous model. Vertical dashed lines indicates the true specified parameter values. The heterogeneous model accurately recovers all parameters, while the homogeneous model underestimates $\Ro$ the NPI parameter (predicting $c_1 \approx 0.24$ instead of the true value $0.3$).}
    \label{fig:Par_Dist_Case_II__b}
\end{figure}

The summary statistics for the estimates of model parameters \( \boldsymbol{\theta_b}\) are shown in Table~\ref{tab:summary_results}, Case II(b), while Figure~\ref{fig:Par_Dist_Case_II__b} shows the parameter distributions and mean estimates. For the homogeneous model, the basic reproduction number \(\Ro\) is underestimated as $2.94$ with $95\%\ \textrm{CI}$ $(2.89,3.00)$,  and \({t_0}\) is estimated as \(14.90\) with $95\%\ \textrm{CI}$ $(14.3,15.5)$. The effect of the NPIs has a mean estimate of \( 0.24\) and $95\%\ \textrm{CI}$ $(0.236,0.243)$, which is biased to enable enough strength to fit the datasets despite the lack of a heterogeneity parameter.

The heterogeneous susceptibility model, on the other hand, accurately estimates the specified parameters as follows: \( {\Ro}\) estimated as \(3.00\) with $95\%\ \textrm{CI}$ $(2.95,3.05)$, \( {\nu}\) estimated as \(1.406\) with $95\%\ \textrm{CI}$ $(1.157,1.768)$, \( {c_1}\) estimated as \(0.299\) with $95\%\ \textrm{CI}$ being $(0.274,0.331)$, and  \( {t_0}\) estimated as \(15.0\) with $95\%\ \textrm{CI}$ $(14.4,15.70)$. Thus both effects of NPIs and heterogeneity are simultaneously estimated accurately. In terms of the AIC, the heterogeneous model indicates a better fit to the data with a score of $713.27$ than the homogeneous model which scored $727.54$.

\begin{figure}[!htb]
    \centering
    \includegraphics[width=1\linewidth]{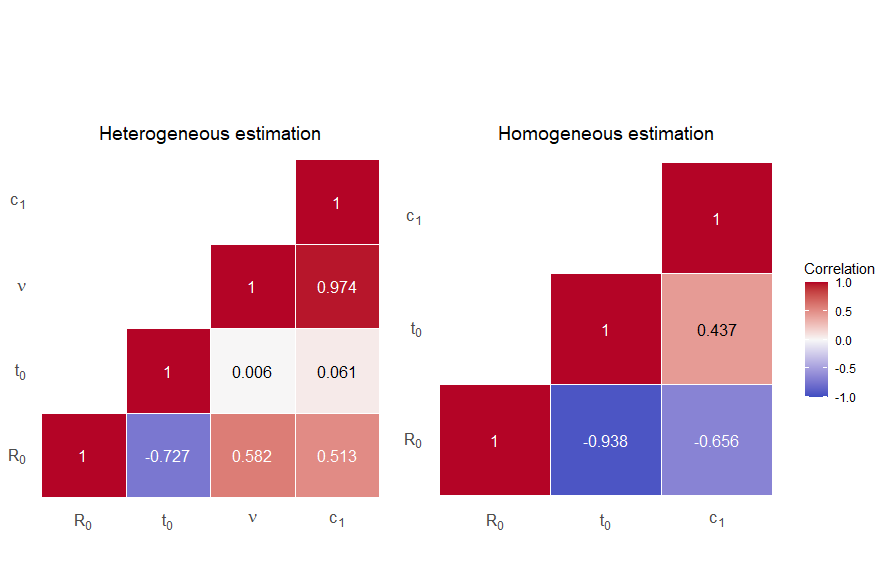}
    \captionsetup{width=0.9\linewidth, justification=justified, aboveskip=5pt}
    \caption{Median parameter correlation heatmaps for Case II(b): heterogeneous data with impact of NPIs. The heterogeneous model (left) shows a very strong positive correlation between $\nu$ and $c_1$ ($0.974$). The homogeneous model (right) shows a very strong negative correlations between $\Ro$ and $t_0$ ($-0.938$).}
    \label{fig:corr_Case_II__b}
\end{figure}

From the correlation plots in Figure~\ref{fig:corr_Case_II__b}, however, some strong correlations are apparent. The homogeneous estimation exhibits negative correlations between the estimates of \(t_0\) and \( \Ro\), and the estimates of \( t_0\) and \(c_1\). For the heterogeneous model, there is a positive correlation between the estimates of \( \Ro\) and \( c_1\), and the estimates of \( \nu \) and \( c_1\), and also a negative correlation between the estimates of \( t_0\) and \( \Ro\). Overall, there are correlations in both model fits but the one that concerns us more here is the correlation of $0.974$ between the impact of the NPIs and CV. This may indicate non-identifiability of the parameters, although from our systematic study we know that, with the heterogeneous model, true values of the parameters have been estimated in all cases. In Supplementary Material Supplementary Section S3 we investigate how these results vary with the initial condition $\Ii(0)$ and intervention strength parameter $c_1$.

To alleviate the correlation between the coefficient of variation \( \nu \) and the impact of NPIs \(c_1\), we will simulate concurrent epidemics, mimicking multiple areas of the same country (as in \citep{Gomes2022} for England and Scotland). Models will be fitted to these data simultaneously by maximizing the sum of the log-likelihoods from the individual epidemic datasets to estimate \(\boldsymbol{\theta_b}\). This will be presented in the following sections.

\begin{figure}[ht]
    \centering
    \includegraphics[trim={0 1.5cm 0 0cm},clip,width=\textwidth]{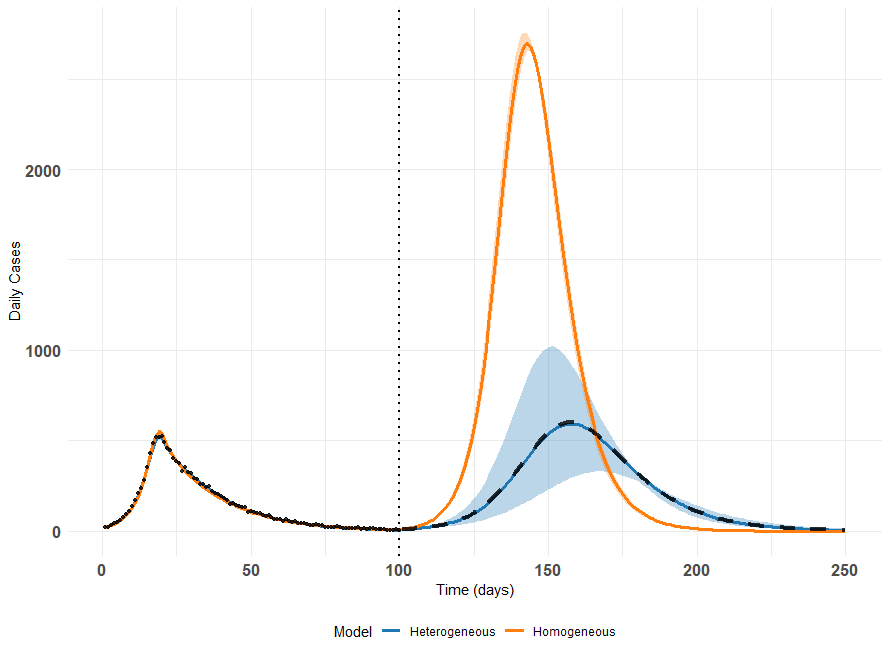}
    \caption{Prediction trajectories for Case II(b): Heterogeneous data with NPIs. Models were fitted to the first 100 days of data (black dots) and then used to forecast the next 150 days. Heterogeneous model fit (blue); homogeneous model fit (orange). The dashed black line represents the ``true'' trajectory generated with the assumed parameter values.}
    \label{fig:case_II_b_prediction}
\end{figure}

Figure~\ref{fig:case_II_b_prediction} concerns predictions. While both models fit the initial 100 days of data reasonably well, their forecasts diverge substantially. The homogeneous model predicts a massive second wave with peak daily cases approaching $2750$ — nearly four times higher than first wave. In contrast, the heterogeneous model predicts a more modest second wave with peak daily cases between $300$ and $1000$, approximately.

\FloatBarrier 

When compared to the ``ground truth'' (dashed black line), it is evident that the heterogeneous model's forecast is remarkably accurate, while the homogeneous model's prediction represents a substantial overestimation. This difference in forecasting performance stems from the homogeneous model's inability to account for the depletion of highly susceptible individuals during the first wave, which naturally limits the severity of subsequent waves. Instead, the homogeneous model attributes the flattening of the curve in the first wave primarily to intervention effects, incorrectly suggesting greater susceptibility of the population for the second wave.

The wider prediction intervals for the heterogeneous model reflect uncertainty in the estimation of CV but, despite this uncertainty, the upper bound of its prediction interval remains well below the homogeneous model's median forecast. This underscores that accounting for population heterogeneity is essential not only for accurate parameter estimation but also for reliable forecasting of epidemic trajectories.


\section{Fitting pairs of epidemics to overcome identifiability issues}

As highlighted in Section~\ref{baseline}, we have encountered correlations between estimated parameters. Given our focus on individual variation in susceptibility, correlations that involve the coefficient of variation ($\nu$) are of special interest (particularly those between $\nu$ with the non-pharmaceutical intervention parameter, $c_1$). 

To address this problem, we draw on a simple yet effective idea. If given a single epidemic, characterised by a set of parameter values and initial conditions, we have difficulties inferring the relative effects of CV and NPIs, we want to ask what might the trajectory look like had the same NPIs started at an earlier of later time point (so that we can observe the effect of CV alone in one realisation while CV+NPIs are both effective in the other). Although we cannot conduct this experiment beyond simulation studies, it is often not difficult in real scenarios to find suitable epidemic trajectories which can paired with our focus epidemic to alleviate identifiability issues. 

This principle was employed by \citet{Gomes2022}, who fitted epidemic models jointly to COVID-19 data from England and Scotland. Here we conduct the first systematic simulation-based investigation of how two-epidemic analyses can resolve identifiability problems inherent to single-epidemic inference exposed in Section~\ref{baseline}. We seek to evaluate: whether analysing simultaneously two epidemics which differ only in epidemic ``age'' at the start of NPIs improves identifiability; and how achieved improvements depend on the strength of NPIs and the difference in epidemic age when NPIs begin. This is conducted by systematically assessing the degree to which estimation precision and parameter correlations are affected.

In the sections that follow, we present detailed profile likelihood and correlation analyses that demonstrate how this approach breaks the confounding between heterogeneity and intervention effects, among other benefits.

\subsection{Eigenvalue and Hessian analysis}

For our parameter vector $\boldsymbol{\theta} = (\Ro, \nu, t_0, c)$, the Hessian matrix at the maximum likelihood estimate (MLE) represents the local curvature of the negative log-likelihood function. To examine parameter identifiability, we examine the structure of the Hessian matrix. The eigen decomposition of this Hessian provides information
\begin{equation}
\mathbf{H} = \mathbf{V} \mathbf{\Lambda} \mathbf{V}^T,
\end{equation}
where $\mathbf{\Lambda} = \text{diag}(\lambda_1, \lambda_2, \ldots, \lambda_p)$ contains eigenvalues ordered by magnitude ($\lambda_1 \geq \lambda_2 \geq \ldots \geq \lambda_p > 0$) and $\mathbf{V} = [\mathbf{v}_1, \mathbf{v}_2, \ldots, \mathbf{v}_p]$ contains the corresponding eigenvectors.

The eigenvalues indicate how quickly the likelihood function changes when moving in different directions in parameter space. Large eigenvalues indicate directions with high curvature, where parameters are well-constrained, while small eigenvalues signal directions with low curvature where parameters are poorly constrained. In addition, the condition number $\kappa = \lambda_1 / \lambda_p$ indicates numerical stability, with large values ($\kappa > 1000$) suggesting practical identifiability issues.

We performed an analysis for synthetic single epidemics, and then pairs of epidemics which share the same parameters except for the timing of NPIs in terms of epidemic age, with a view to alleviating the correlation as we shall see shortly.
In our analysis of the Hessian for single-epidemic fits, we consistently observe one eigenvalue significantly smaller than the others, indicating a direction in parameter space with low curvature and thus poor identifiability with a large condition number. However, if two epidemics curves are simultaneously used to compute the Hessian, the condition number can be substantially reduced.

The correlation matrix derived from the inverse Hessian,
\begin{equation}
\text{Corr}_{ij} = \frac{(\mathbf{H}^{-1})_{ij}}{\sqrt{(\mathbf{H}^{-1})_{ii} (\mathbf{H}^{-1})_{jj}}},
\end{equation}
quantifies the statistical dependencies between parameter estimates.

\subsection{Profile likelihood analysis}

The profile likelihood approach offers a practical method to visualize identifiability issues by exploring how the likelihood changes when one parameter is varied while all others are optimized. For a parameter $\theta_i$, the profile likelihood is defined as
\begin{equation}
PL(\theta_i) = \max_{\theta_j, j \neq i} L(\boldsymbol{\theta}).
\end{equation}
This approach involves fixing one parameter at various values within a plausible range and re-optimizing all other parameters through maximum likelihood estimation. This process generates a curve of maximum log-likelihood values as a function of the fixed parameter, enabling the construction of confidence intervals and revealing structural correlations between parameters.

We applied this method to 200 simulated datasets using the reduced SEIR model with gamma-distributed susceptibility. Each dataset was generated with parameters $\Ro=3.0$ (basic reproduction number), $\nu=1.414$ (coefficient of variation), $t_0=15$ days (behavioural change), and $c_1=0.3$ ($1-$ intervention strength), then fitted using maximum likelihood estimation. For the two-epidemic scenario, we simultaneously simulated a ``baseline'' epidemic (initial conditions $\Ei(0)=100$, $\Ii(0)=40$) and an ``auxiliary'' epidemic ($\Ei(0)=1000$, $\Ii(0)=400$) with identical parameters. The different initial conditions account for the difference in epidemic ages at each time point. 
For each dataset, we profiled all four  parameters and analysed the resulting likelihood profiles, parameter correlations, confidence intervals, and eigenvalue structures of the Hessian matrices. During optimization, we employed parameter transformations (logarithmic for $\Ro$, $\nu$, and $t_0$; logit transformation for $c_1$) to ensure that parameter constraints were maintained during optimization. Confidence intervals were determined using the chi-square cutoff method, where parameter values yielding likelihood ratio statistics below the threshold defined by $\chi^2_{0.95,1} = 3.84$ were included in the $95\%$ confidence interval. 



Figures S9-S12, in Supplementary Material Section S4.1, show representative profile likelihood curves for all four parameters from a randomly selected sample dataset, comparing the single-epidemic versus two-epidemic approaches.
Table~\ref{tab:estimates_comparison} presents the parameter estimation results across all 200 datasets, comparing the single and two-epidemic approaches. The most substantial improvement is observed for the coefficient of variation parameter ($\nu$), with a CI width reduction of approximately $93\%$. The $86\%$ reduction in CI width for the intervention strength parameter ($c_1$) is also noteworthy. These width-reduction improvements can be visualised in Supplementary Figure S13 (Supplementary Material Section S4.2) which displays the $95\%$ confidence intervals of parameter estimates from all the datasets, along with the respective median values.



\begin{table}[htbp]
\begin{threeparttable}
\centering
\caption{Parameter estimates and confidence interval comparisons ($c_1 = 0.3$).}
\begin{tabular}{lccccc}
\toprule
\multirow{2}{*}{Parameter} & \multicolumn{2}{c}{Single epidemic} & \multicolumn{2}{c}{Two epidemics} & \multirow{2}{*}{CI width reduction} \\
\cmidrule(lr){2-3} \cmidrule(lr){4-5}
& Mean (SD) & CI width & Mean (SD) & CI width & \\
\midrule
$R_0$ & 3.00 (0.03) & 0.090 & 3.00 (0.01) & 0.038 & 57.6\% \\
$\nu$ & 1.42 (0.18) & 0.571 & 1.41 (0.01) & 0.043 & 92.5\% \\
$t_0$ & 14.98 (0.36) & 1.174 & 15.01 (0.19) & 0.667 & 43.2\% \\
$c_1$ & 0.30 (0.02) & 0.055 & 0.30 (0.00) & 0.008 & 86.1\% \\
\bottomrule
\end{tabular}
\label{tab:estimates_comparison}
\end{threeparttable}
\end{table}

To further examine the statistical performance of both approaches, we analysed the relative bias and coverage probabilities of the confidence intervals across all 200 datasets, as shown in Supplementary Tables S1 and S2. The results demonstrate that the two-epidemic approach not only provides more precise estimates but also exhibits lower relative bias for most parameters. 

Finally, a numerical stability analysis conducted in Supplementary Material Section S4.3 concludes that the two-epidemic approach produces a better-conditioned optimization problem, with lower and less variable condition numbers in comparison with the single-epidemic problem. 






\subsection{Parameter correlation analysis}

\begin{figure}[!htb]
    \centering
    \includegraphics[width=0.9\textwidth]{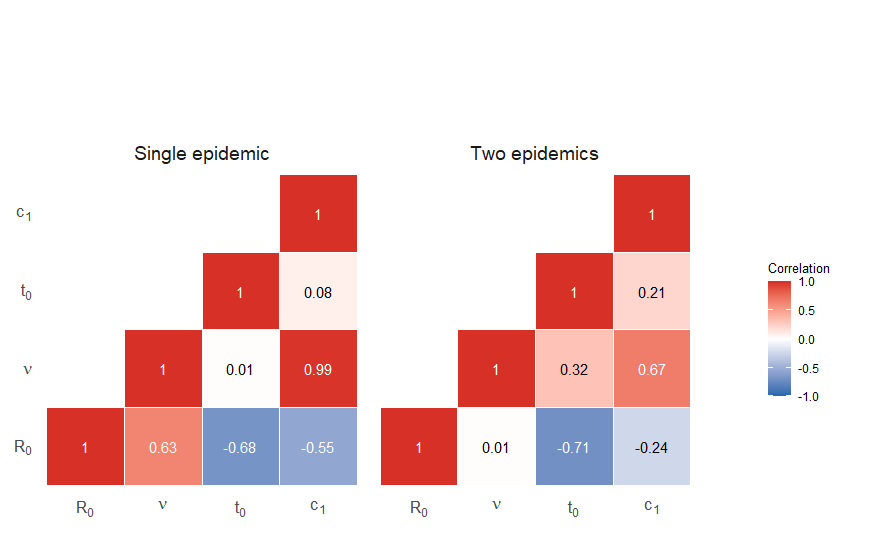}
    \caption{Median parameter correlation heatmaps comparing single-epidemic (left) and two-epidemic (right) approaches. Lower correlations in the two-epidemic approach indicate improved parameter identifiability.}    \label{fig:correlation_heatmap_comparison}
\end{figure}

We computed pairwise parameter correlations between the 200 estimates to understand how the two-epidemic approach affects parameter identifiability. Figure~\ref{fig:correlation_heatmap_comparison} shows the correlation matrices for both approaches.

While some parameter pairs show reduced correlations, others maintain similar levels, reflecting the complex trade-offs in parameter identifiability. Most importantly, the two-epidemic approach successfully reduces correlations for the parameter pairs that concern this study the most - those that directly impact our ability to separate heterogeneity effects from intervention effects. This provides the mechanistic foundation for the improved precision and identifiability observed throughout this analysis.

\subsection{Sensitivity Analysis} \label{subsec:sensitivity_anal}

Sensitivity analysis examines how changes in model parameters affect the model output. For a parameter $\theta_j$, the sensitivity function $S_{ij}(t)$ measures how the observable data at time $t$ changes with respect to that parameter:
\begin{equation} \label{sensitivity}
S_{ij}(t) = \frac{\partial y_i(t)}{\partial \theta_j}
\end{equation}
where $y_i(t)$ is the observable (daily incidence) for epidemic $i$ at time $t$. 
Intuitively, when two parameters have similar sensitivity patterns, compensation effects are created, such that changes in one parameter can be offset by changes in another to produce similar epidemic curves. However, when sensitivity patterns are different (orthogonal), the parameters become more distinguishable. 

We conducted a comparative sensitivity analysis for single and two-epidemic approaches. As previously, we simulated epidemic datasets using parameter values $\Ro = 3.0$, $\nu = 1.414$, $t_0 = 15$ days, $c_1 = 0.3$ (initial conditions, $\Ei(0)=100$, $\Ii(0)=40$ for the focal epidemic, and $\Ei(0)=1000$, $\Ii(0)=400$ for the auxiliary). We then
performed local sensitivity analyses \eqref{sensitivity}, using finite differences to approximate derivatives of model outputs with respect to parameters \citep{press2007numerical}. 

Figure S15, in Supplementary Material Section S5, illustrates the reduction in parameter compensation as we move from single to two-epidemic approaches. For single epidemics, the sensitivity patterns for CV ($\nu$) and intervention strength ($c_1$) follow nearly identical temporal trajectories, creating identifiability challenges. In contrast, the two-epidemic approach produces orthogonal sensitivity patterns, effectively breaking compensation mechanisms, especially around a time window where natural epidemic dynamics begin to be disrupted by behavioural change and NPIs, since this occurs at different phases for different epidemics. These tests provide further empirical support for the adoption of two-epidemic frameworks as a means to alleviate parameter identifiability issues.

\section{Sensitivity to initial conditions of auxiliary epidemic}

We now examine how the initial conditions of the auxiliary epidemic affect the gain in parameter identifiability relative to the single-epidemic approach. We present results for three values of the intervention parameter ($c_1 = 0.2, 0.3, 0.4$). All other parameters remain fixed as usual: $\Ro = 3.0$, $\nu = 1.414$, and $t_0 = 15$ days.

Specifically, we simulated two concurrent epidemics: a focal epidemic with $\Ii(0) = 40$;  and an auxiliary epidemic with $\Ii(0) \in \{ 20,40,80, 160, 320, 400\}$. For each scenario, we generated 200 synthetic datasets and analysed parameter correlations using maximum likelihood estimation.

\begin{figure}[!htb]
    \centering
    \includegraphics[width=0.95\textwidth]{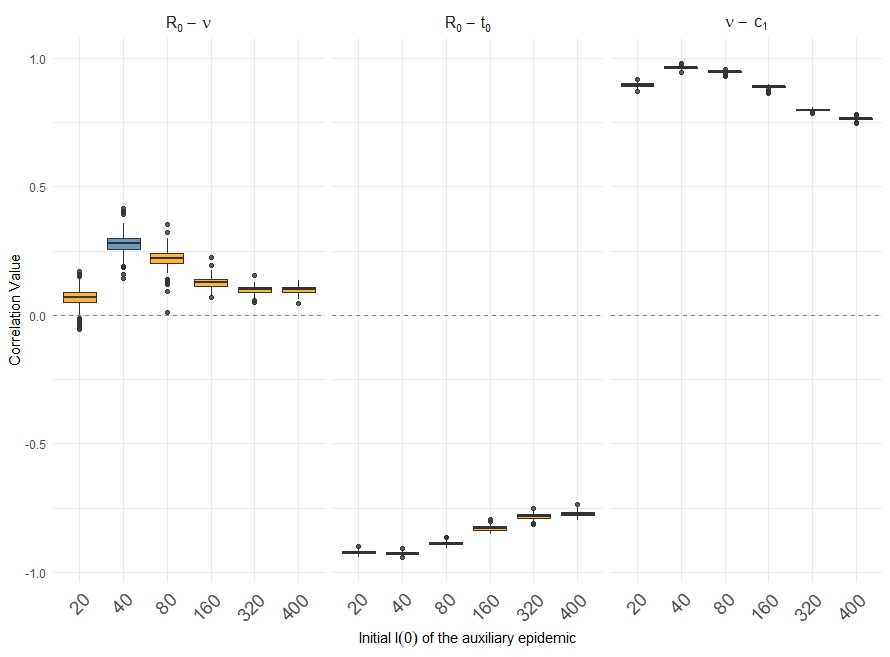}
    \caption{Correlation distributions in terms of the initial condition of the auxiliary epidemic ($\Ii(0) \in \{20,40, 80, 160, 320, 400\}$) for intervention parameter $c_1 = 0.4$. The baseline epidemic is initiated with $\Ii(0)=40$ in all cases. Cases where baseline and auxiliary epidemics coincide (i.e., have the same initial conditions) are marked in blue. }
    \label{fig:corr_dist_04}
\end{figure}

\begin{figure}[!htb]
    \centering
    \includegraphics[width=0.95\textwidth]{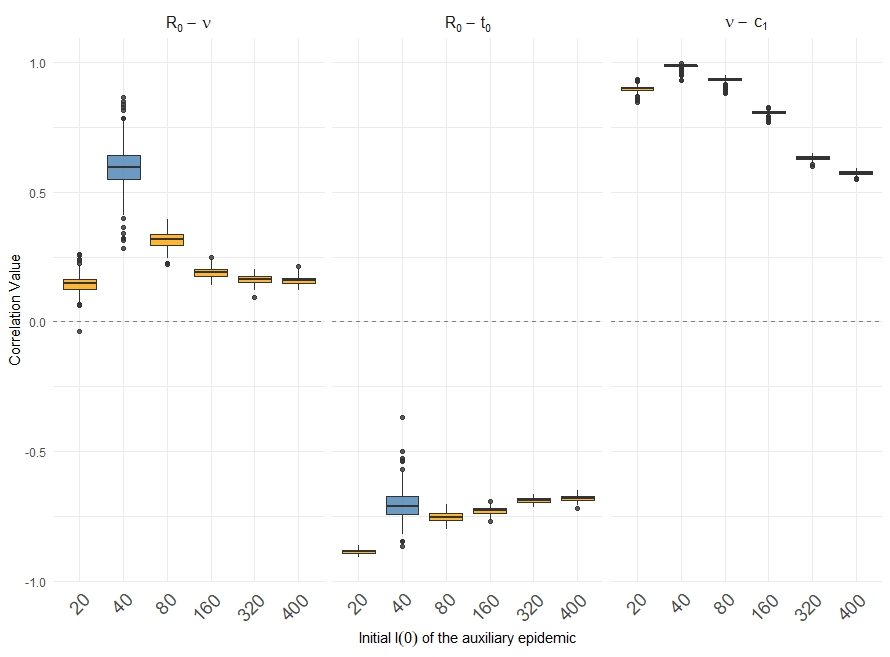}
    \caption{Correlation distributions in terms of the initial condition of the auxiliary epidemic ($\Ii(0) \in \{20,40, 80, 160, 320, 400\}$) for intervention parameter $c_1 = 0.3$. The baseline epidemic is initiated with $\Ii(0)=40$ in all cases. Cases where baseline and auxiliary epidemics coincide (i.e., have the same initial conditions) are marked in blue. }
    \label{fig:corr_dist_03}
\end{figure}

\begin{figure}[!htb]
    \centering
    \includegraphics[width=0.95\textwidth]{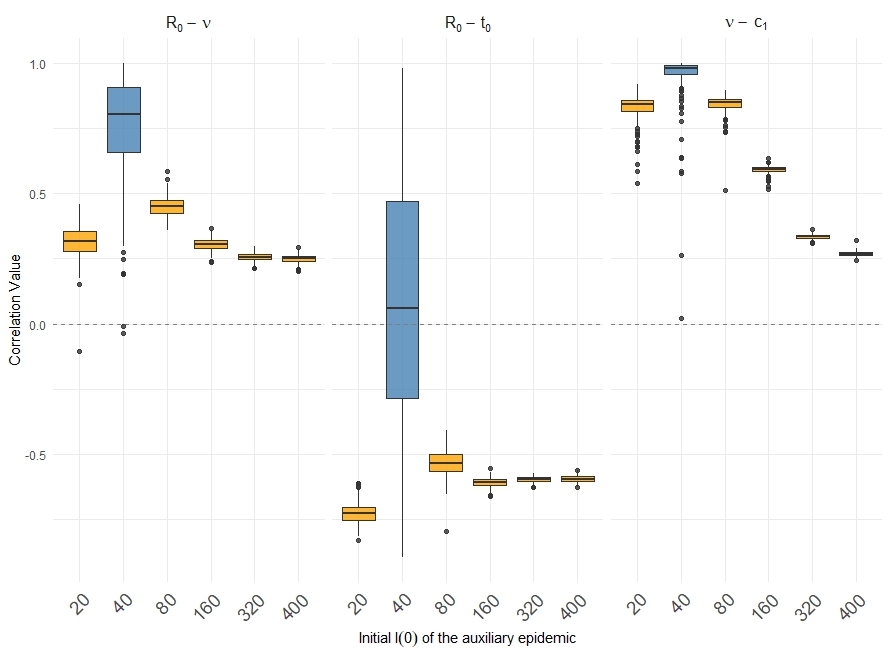}
    \caption{Correlation distributions in terms of the initial condition of the auxiliary epidemic ($\Ii(0) \in \{ 20,40,80, 160, 320, 400\}$) for intervention parameter $c_1 = 0.2$. The baseline epidemic is initiated with $\Ii(0)=40$ in all cases. Cases where baseline and auxiliary epidemics coincide (i.e., have the same initial conditions) are marked in blue. }
    \label{fig:corr_dist_02}
\end{figure}

Figures~\ref{fig:corr_dist_04}, \ref{fig:corr_dist_03}, and \ref{fig:corr_dist_02} show how parameter correlations vary with the initial condition of the auxiliary epidemic.

As expected, all correlations tend to decrease as the initial condition of the auxiliary epidemic deviates from that of the focal (baseline). In addition, we see that as the intervention becomes stronger (decreasing $c_1$), the correlation between $\Ro$ and $\nu$ increases, while those between $\Ro$ and $t_0$ and between $\nu$ and $c_1$ decrease. This effect is especially pronounced in the case of $\nu$-$c_1$. In the COVID-19 study of \cite{Gomes2022}, $c_1$ was estimated between $0.2$ and $0.3$, in agreement with the empirical study of \cite{Jarvis2020}, which happens to be favourable regime from the identifiability perspective.

\section{Conclusions}

In this study, we conducted the first systematic investigation of parameter identifiability for the SEIR model with inter-individual variation in susceptibility to infection. The work consisted in simulating datasets using a stochastic version of the model with strategically chosen parameter values - with or without heterogeneity, with or without non-pharmaceutical interventions - and then conducting maximum likelihood estimation to assess how well we could retrieve the assumed parameter values. The key findings are as follows:
\begin{itemize}
    \item Model parameters are generally estimated accurately, but there is a tendency for high uncertainty in the estimated CV of susceptibility when fitting to data that included NPIs.
    \item Strong correlations among pairs of estimated parameters are generally apparent, whether heterogeneity is included as a parameter or not, indicating potential issues with identifiability.
    \item The issues highlighted above can be substantially alleviated by simultaneously fitting two epidemic trajectories with some shared parameters.
\end{itemize}
The strategy of fitting multiple epidemics with shared parameters to gain identifiability has been used successfully in real-world investigations (e.g., \cite{White2007,Aguas2008,Gomes2022}), but a systematic study of the strengths of the approach was, to our knowledge, lacking. We hope that the significance of the results reported here will motivate others to conduct such investigation for other systems, and apply the approach in real settings.


\section*{Competing interests}
No competing interest is declared.

\section*{Author contributions statement}
All authors conceived the study, implemented the models, analysed the results and wrote the paper.

\section*{Data and code availability}
Computer code used to simulate epidemics and estimate model parameters is available from https://github.com/ibrahimStrathclyde/On-the-simultaneous--inference--of--epidemic-susceptibility-distributions-and-NPI

\section*{Acknowledgments}
M.G.M.G. is partially funded by FCT – Fundação para a Ciência e a Tecnologia, I.P., projects UIDB/00297/2020 (https://doi.org/10.54499/UIDB/00297/2020) and UIDP/00297/2020 (https://doi.org/10.54499/UIDP/00297/2020) (Center for Mathematics and Applications).
I.M. is funded by Petroleum Development Fund (PTDF), Nigeria.
\appendix
\section{Supplementary material}
\label{app1}

Supplementary material related to this article can be found online
at...



\bibliographystyle{elsarticle-harv} 
\bibliography{reference}







\end{document}


\maketitle



\section{Setting initial conditions for simulations}

To set the initial conditions for simulations in Section 3 (``Simulations''), we assume a number of infectious individuals at time zero (i.e., $\Ii(0)$) and, from that, approximate how many are expected to be in the other model compartments. Making the approximations that no-one has recovered from infection by that time ($\Ri(0)=0$), and that depletion of susceptibles and behavioural change have not impacted early infection dynamics, system (4) of the main text is approximated by the linear system
\begin{equation}\label{Linear_system}
\begin{split}
 \frac{\dd \Ei}{\dd t}&= \beta\ (\rho\ \Ei+\Ii)\ - \delta\ \Ei, \\
\frac{\dd \Ii}{\dd t}&= \delta\ \Ei-\gamma\ \Ii,
\end{split}
\end{equation}
and $\Si(0)=\Ni-\Ei(0)-\Ii(0)$. Then we use system \eqref{Linear_system} to determine $\Ei(0)$. In matrix form, the system becomes
\begin{equation}\label{matrix}
 \frac{\dd}{\dd t}\left(\begin{array}{c}\Ei\\ \Ii\end{array}\right) = \underbrace{\left(\begin{array}{cc}\rho\beta-\delta& \beta\\ \delta& -\gamma\end{array}\right)}_A \left(\begin{array}{c}\Ei\\ \Ii\end{array}\right).
\end{equation}
The eigenvalues of $A$ are
\begin{equation}\label{eigen_values}
 \lambda_{\pm}= \frac{\rho\beta-\delta-\gamma\pm \sqrt{(\rho\beta-\delta+\gamma)^2}+4\delta\beta}{2}.
\end{equation}
Substituting the values of the parameters in Table 1 of the main text (namely, $\Ni=100,000$ individuals, $\delta=1/5.5$ per day, $\gamma=1/4$ per day, $\rho=0.5$, and $\Ro=3$), and using the relation $\Ro= \beta\left(\frac{\rho}{\delta}+\frac{1}{\gamma}\right)$ (main text, formula (3)), we see that $\lambda_+\approx 0.2>0$ and $\lambda_-\approx -0.09<0$. Hence, solutions of system \eqref{Linear_system} tend asymptotically to the eigenvector corresponding to eigenvalue $\lambda_+$.

\begin{figure}[hbt!]
    \centering
        \includegraphics[trim={0 7.5cm 0 7.5cm},clip,width=0.6\textwidth]{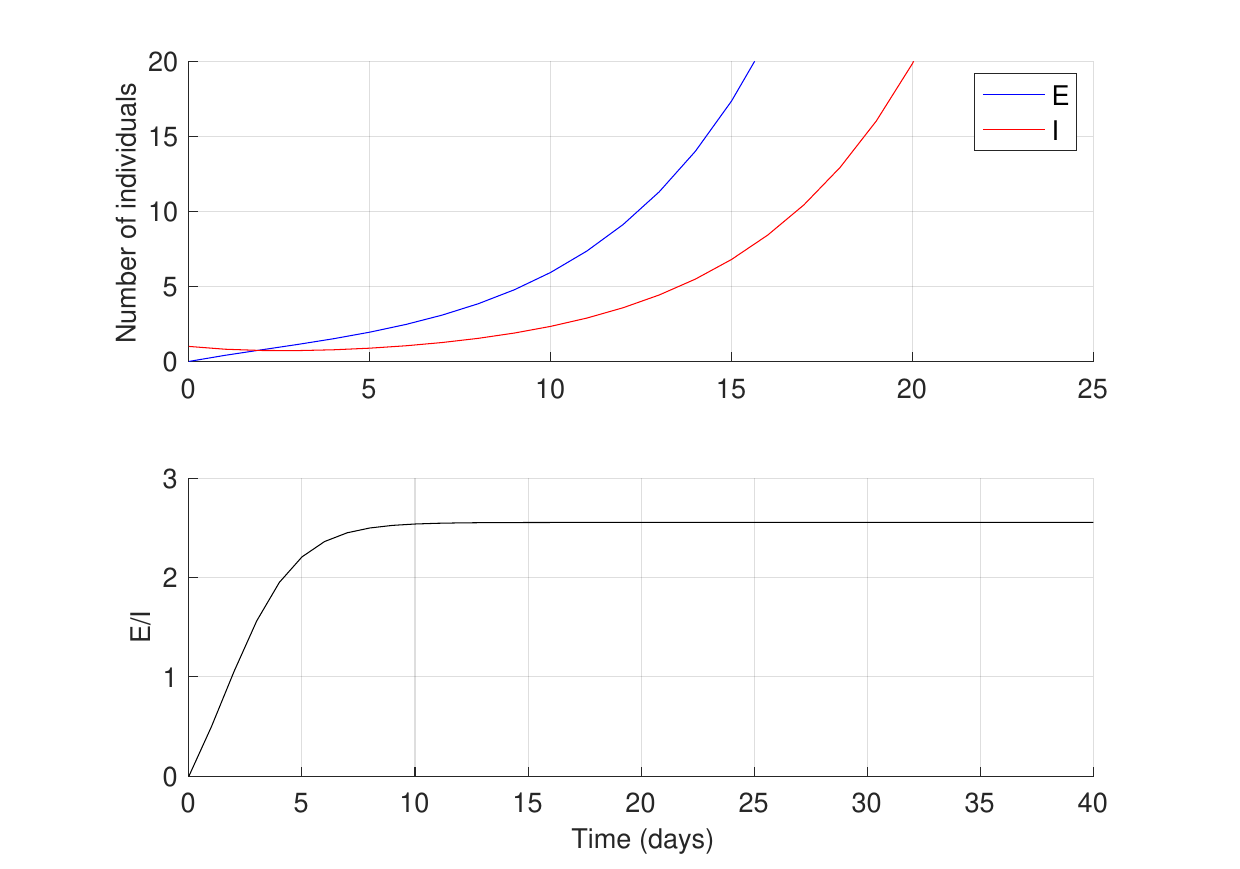}
    \caption{Solution of linear system \eqref{Linear_system} starting from $(\Ei,\Ii)=(0,1)$. Trajectories of $\Ei$ and $\Ii$ over time (top); ratio $\Ei/\Ii$ (bottom).}
   \label{fig:ICs}
\end{figure}

The eigenvectors are $(\Ei_\pm\quad \Ii_\pm)^T$ such that
\begin{equation}\label{eigen_vectors}
 \Ei_\pm= \frac{\gamma+\frac{1}{2}(\rho\beta-\delta-\gamma)\pm\frac{1}{2}\sqrt{(\rho\beta-\delta+\gamma)^2+4\delta\beta}}{\delta}\Ii_\pm,
\end{equation}
and hence solutions of system \eqref{Linear_system} take the form
\begin{equation}\label{soln}
\begin{split}
 \left(\begin{array}{c}\Ei(t)\\ \Ii(t)\end{array}\right) &= c_+\left( \begin{array}{c}\gamma+\frac{1}{2}(\rho\beta-\delta-\gamma)+\frac{1}{2}\sqrt{(\rho\beta-\delta+\gamma)^2+4\delta\beta}\\ \delta\end{array}\right) e^{\lambda_+t} \\ &+ c_-\left( \begin{array}{c}\gamma+\frac{1}{2}(\rho\beta-\delta-\gamma)-\frac{1}{2}\sqrt{(\rho\beta-\delta+\gamma)^2+4\delta\beta}\\ \delta\end{array}\right) e^{\lambda_-t}
 \end{split}
\end{equation}
Since $\lambda_-<0$, all solutions tend to the first term on the right hand side of \eqref{soln}, which satisfies a ratio $\Ei(t)/\Ii(t)$ of approximately $2.55$. Moreover, the convergence is sufficiently fast for this to be an excellent approximation for all scenarios simulated in this paper. Specifically, all our simulations are conducted for $\Ii(0)\geq 20$ and Figure \ref{fig:ICs} shows that the ratio $\Ei/\Ii$ is already very close to $2.5$ well before $\Ii$ reaches $20$.

\FloatBarrier 

\section{Model fittings}

Here we show a sample of the fittings associated with results reported in Section 5 (``Baseline analysis'') of the main text.

\begin{figure}[!htb]
    \centering
    \begin{subfigure}[t]{0.85\textwidth}
        \centering
        \includegraphics[trim={0 1.5cm 0 0cm},clip,width=\textwidth]{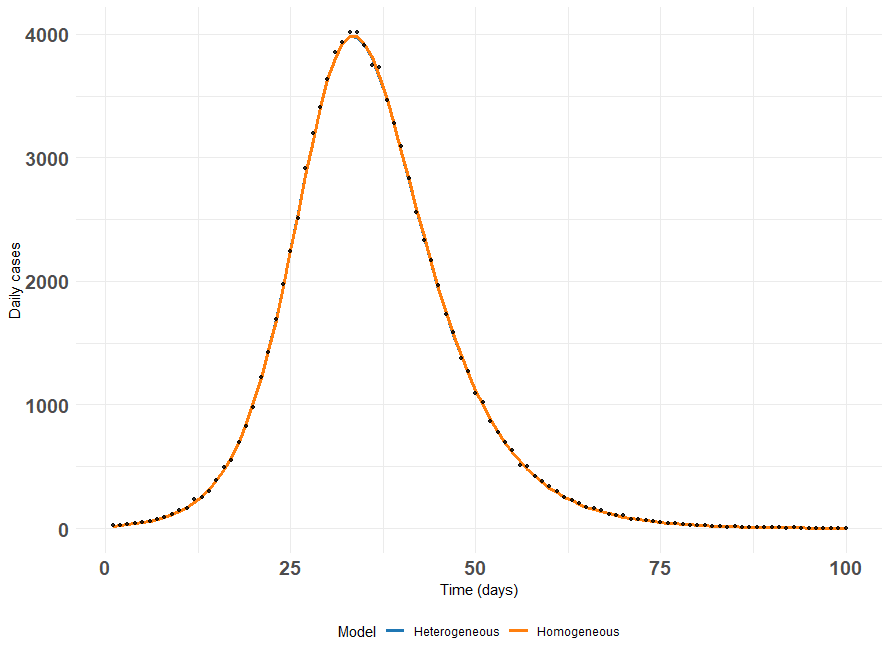}
    \end{subfigure}
    \begin{subfigure}[t]{0.85\textwidth}
        \centering
        \includegraphics[trim={0 1.5cm 0 0cm},clip,width=\textwidth]{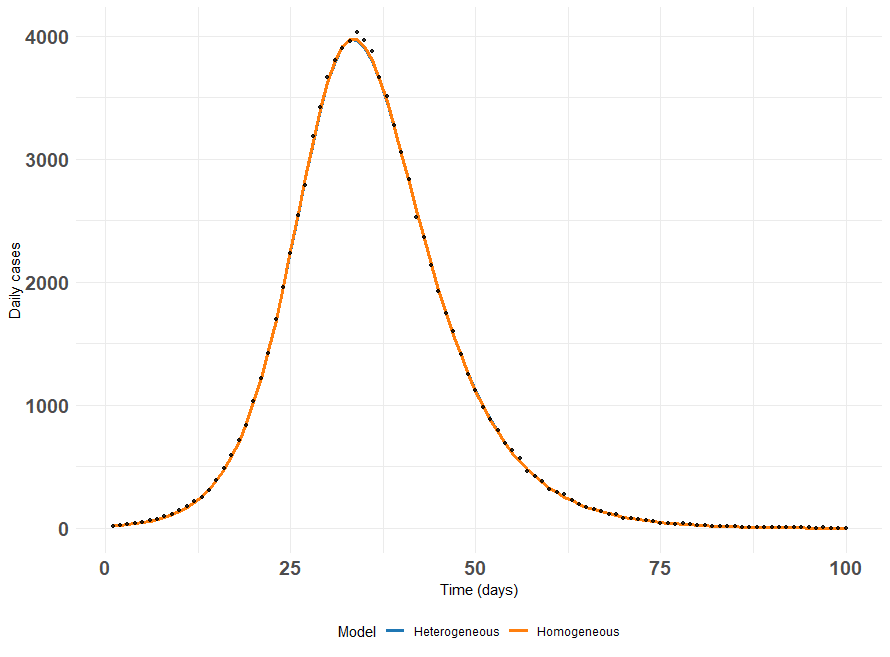}
    \end{subfigure}    
    \caption{Case I(a): Homogeneous data without NPIs. Black dots represent simulated daily cases. Heterogeneous model fits (blue lines) and homogeneous model fits (orange lines) are practically coincident. Fittings conducted without NPIs (scenario (i), top); fittings allowing NPI effects (scenario (ii), bottom).}
    \label{fig:fits_homdata_with_NPIs}
\end{figure}
\FloatBarrier 

\begin{figure}[!htb]
    \centering
    \includegraphics[trim={0 1.5cm 0 0cm},clip,width=0.85\textwidth]{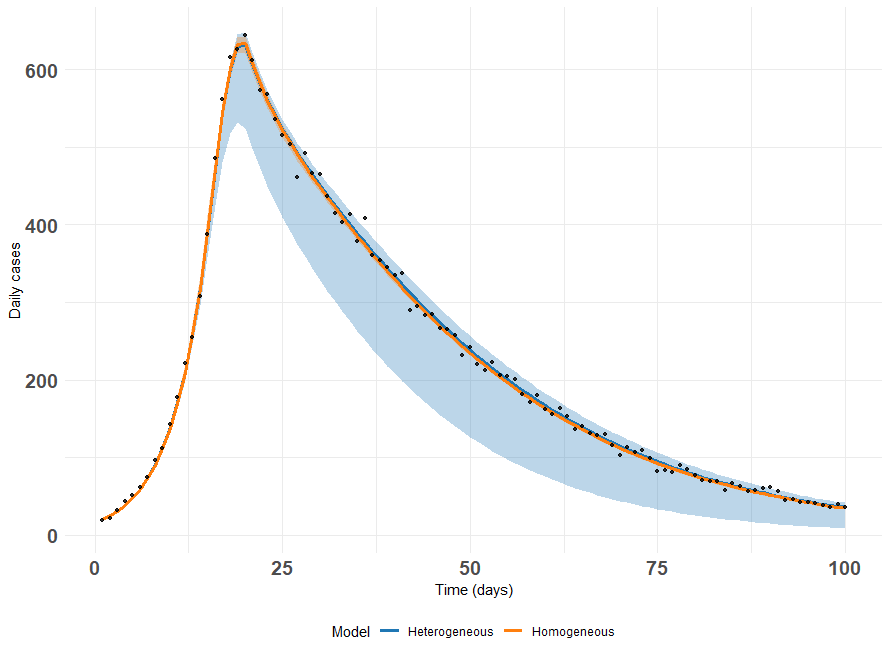}
    \caption{Case I(b): Homogeneous data with NPIs. Black dots represent simulated daily cases. Heterogeneous model fit (blue line) and homogeneous model fit (orange line) are practically coincident. Fittings allowing NPI effects.}
    \label{fig:case_I_b}
\end{figure}
\FloatBarrier 

\begin{figure}[!htb]
    \centering
    \begin{subfigure}[t]{0.85\textwidth}
        \centering         
        \includegraphics[trim={0 1.5cm 0 0cm},clip,width=\textwidth]{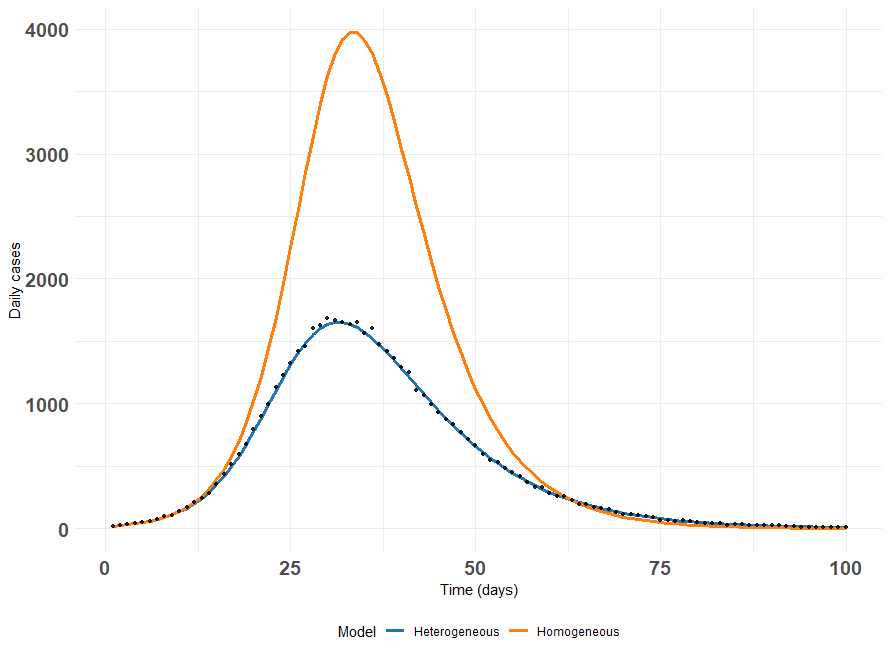}
    \end{subfigure}   
    \begin{subfigure}[t]{0.85\textwidth}
        \centering
\includegraphics[trim={0 1.5cm 0 0cm},clip,width=\textwidth]{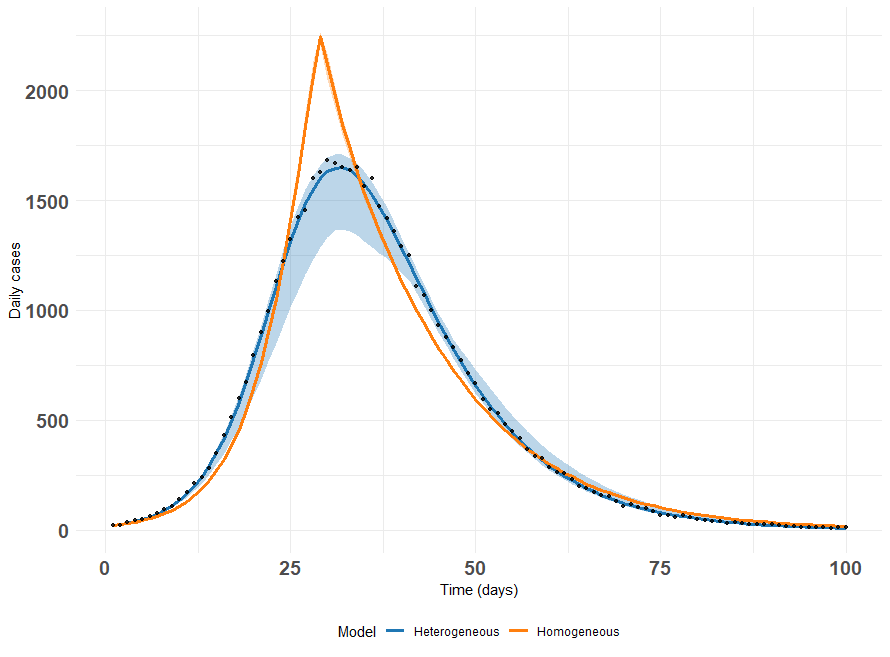}   
    \end{subfigure}   
    \caption{Case II(a): Heterogeneous data without NPIs.  Black dots represent simulated daily cases. Heterogeneous model fits (blue lines); homogeneous model fits (orange lines). Fittings conducted without NPIs (scenario (i), top); fittings allowing NPI effects (scenario (ii), bottom).}
    \label{fig:case_II_a}
\end{figure}
\FloatBarrier 

\begin{figure}[!htb]
    \centering
    \includegraphics[width=0.85\textwidth]{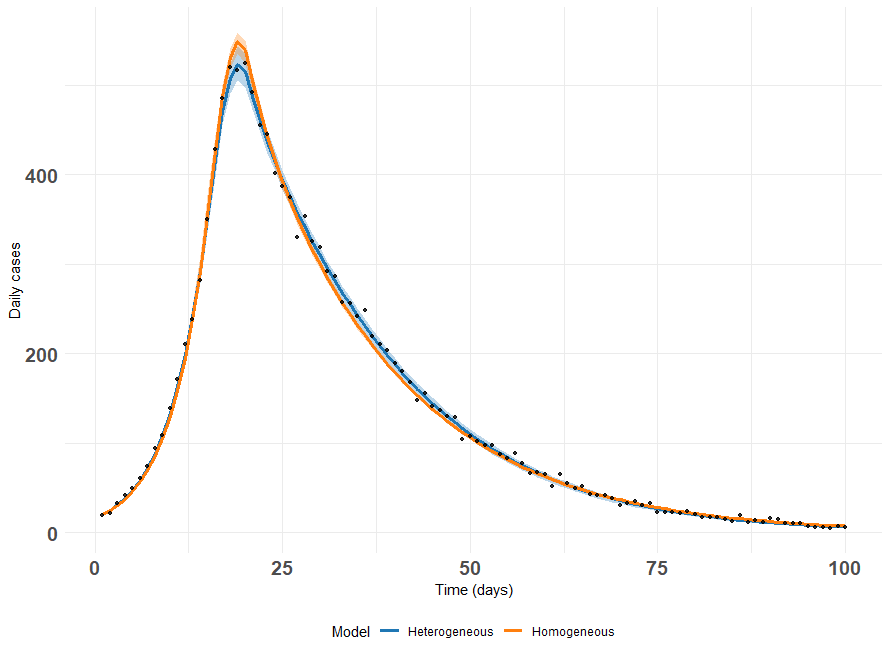}
    \caption{Case II(b): Heterogeneous data with NPIs. Black dots represent simulated daily cases. Heterogeneous model fit (blue line); homogeneous model fit (orange line). Fittings allowing NPI effects.}
    \label{fig:case_II_b}
\end{figure}
\FloatBarrier 

\section{Sensitivity to initial conditions and intervention strength} \label{sec:sensitivity1}

Here we investigate the sensitivity of parameter identifiability to initial conditions and intervention strengths. We examine how the size of the initial infected population ($\Ii(0)$) and the parameter controlling the maximum strength of the intervention ($c_1$) affect the correlation structure between model parameters.

We conducted a systematic analysis using simulated epidemic data with varying initial conditions and intervention strength. For each initial infected population
$\Ii(0) \in \{20,40, 80, 160, 320, 400\}$, we generated 200 synthetic epidemic datasets using the heterogeneous SEIR model with fixed parameters ($\Ro=3$, $\nu=1.414$, $t_0=15$ days). We examined three intervention strengths $c_1 \in \{0.2, 0.3, 0.4\}$. We estimated parameters for each scenario.

\begin{figure}[!htb]
    \centering
    \includegraphics[width=0.95\textwidth]{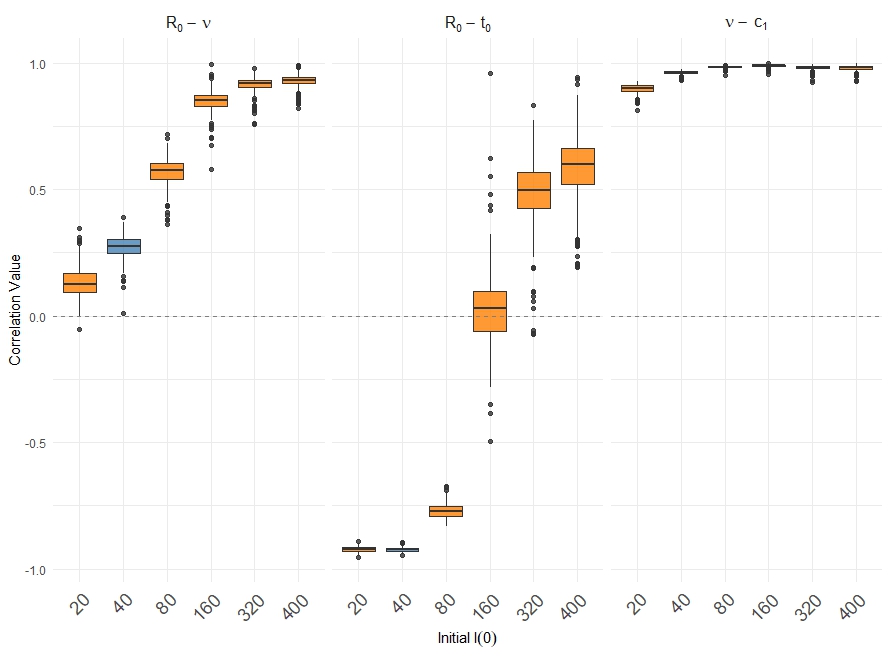}
    \caption{Correlation distributions in terms of initial condition ($\Ii(0) \in \{20,40, 80, 160, 320, 400\}$) for intervention parameter $c_1 = 0.4$. Cases with initial conditions used in the baseline analysis ($\Ii(0)=40$) are marked in blue. }
    \label{fig:corr_dist_04_single}
\end{figure}

\begin{figure}[!htb]
    \centering
    \includegraphics[width=0.95\textwidth]{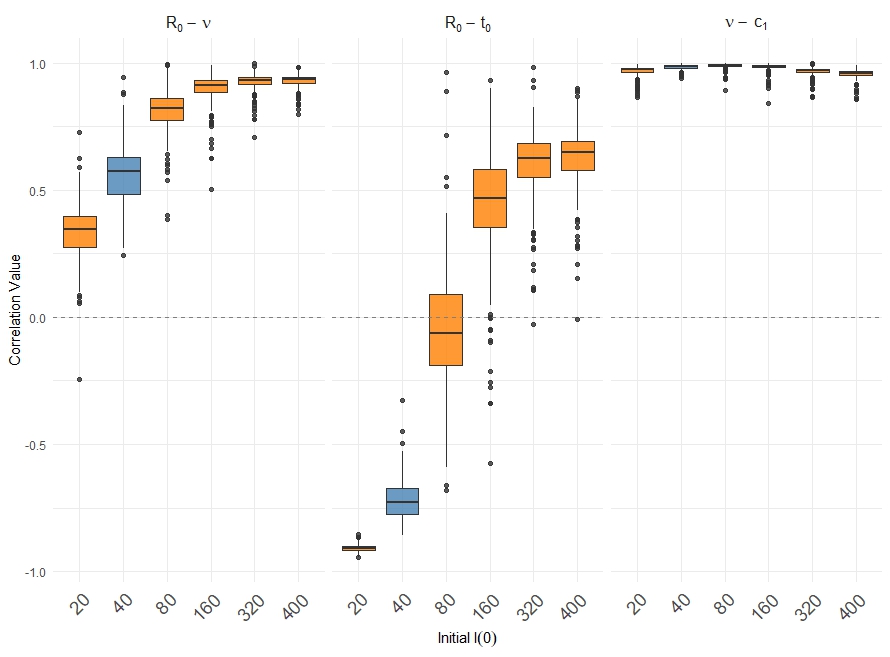}
    \caption{Correlation distributions in terms of initial condition ($\Ii(0) \in \{40, 80, 160, 320, 400\}$) for intervention parameter $c_1 = 0.3$. Cases with initial conditions used in the baseline analysis ($\Ii(0)=40$) are marked in blue. }
    \label{fig:corr_dist_03_single}
\end{figure}

\begin{figure}[!htb]
    \centering
    \includegraphics[width=0.95\textwidth]{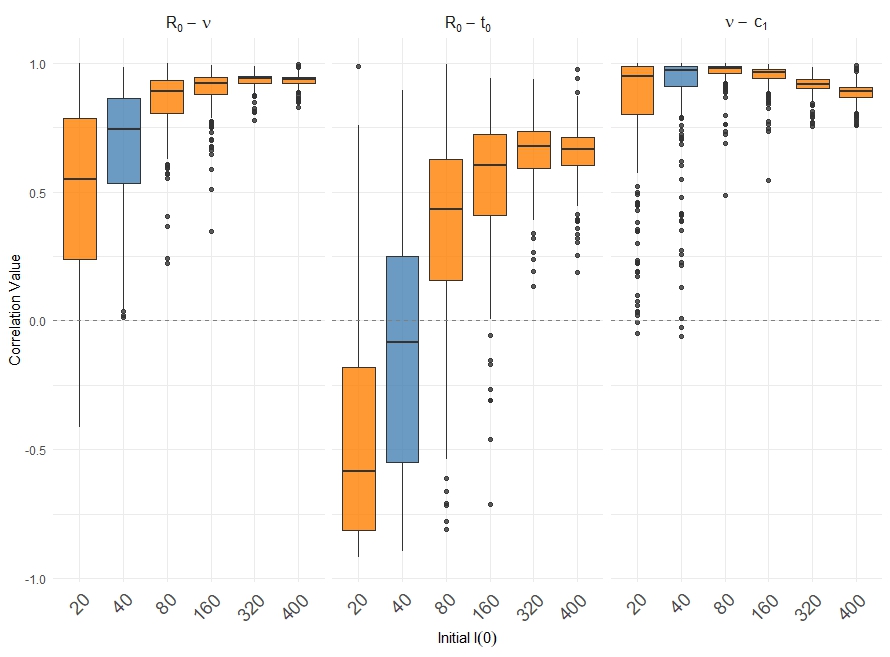}
    \caption{Correlation distributions in terms of initial condition ($\Ii(0) \in \{20,40, 80, 160, 320, 400\}$) for intervention parameter $c_1 = 0.2$. Cases with initial conditions used in the baseline analysis ($\Ii(0)=40$) are marked in blue. }
    \label{fig:corr_dist_02_single}
\end{figure}

Figures~\ref{fig:corr_dist_04_single}, \ref{fig:corr_dist_03_single}, and \ref{fig:corr_dist_02_single} show the distribution of parameter correlations across initial conditions for both single and two epidemic approaches.

\FloatBarrier 

\FloatBarrier 

\FloatBarrier 

Correlations between $\Ro$ and $\nu$ increase sharply with the initial infected population, highlighting the importance of information provided by the initial growth phase for parameter identification. Correlations between $\nu$ and $c_1$ are high across the entire range of initial conditions. Correlations between $\Ro$ and $t_0$ are more difficult to synthesise. They are negative for small initial infected conditions and positive when the initial numbers infected are higher. Since both increasing $\Ro$ and $t_0$ increase growth, we might expect the correlation to be negative. An explanation for how it becomes positive when initial conditions are higher does not appear to be simple and it may involve more complex interactions where other parameters may need to be invoked. 






\section{Profile likelihood analysis}

Here we provide materials supplementing Section 6 (``Profile likelihood analysis and parameter identifiability'') of the main text.

\subsection{Profile likelihood curves}

\begin{figure}[!htb]
    \centering
    \begin{subfigure}[t]{0.8\textwidth}
        \centering
        \includegraphics[width=\textwidth]{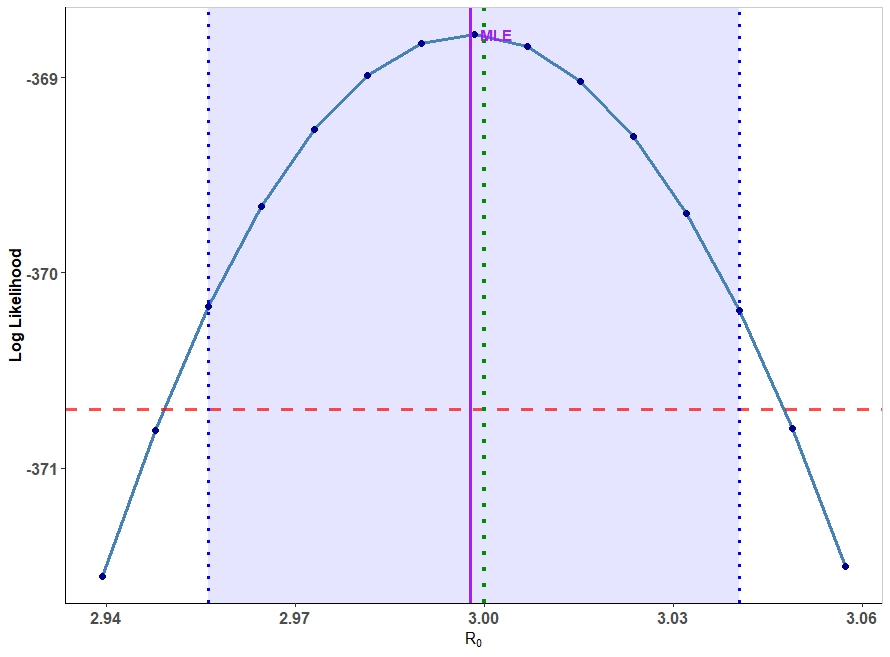}
    \end{subfigure}    
    \begin{subfigure}[t]{0.8\textwidth}
        \centering
        \includegraphics[width=\textwidth]{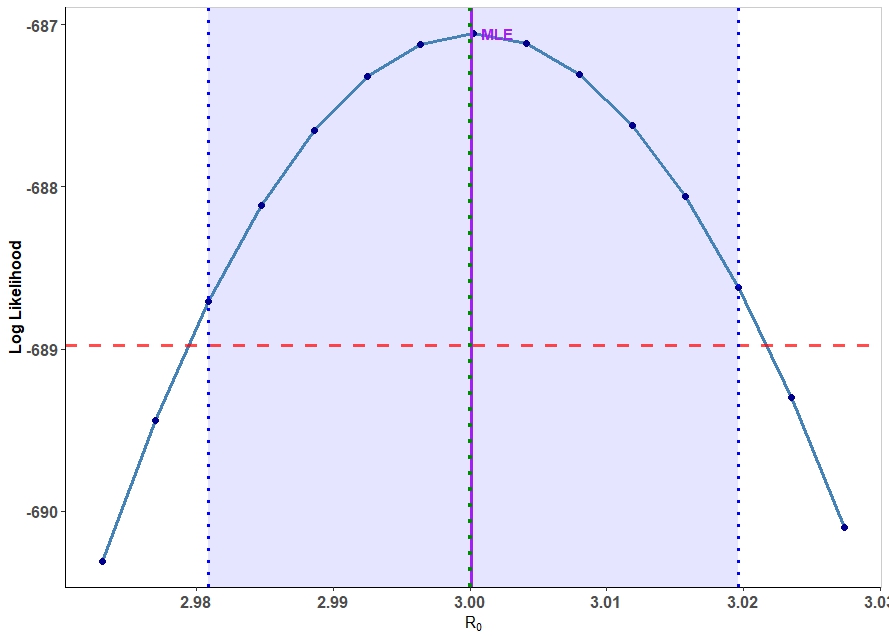}
    \end{subfigure}    
    \caption{Profile likelihood curves for the basic reproduction number $\mathcal{R}_0$ using single-epidemic (top) and two-epidemic (bottom) models. Horizontal dashed red lines represent the $95\%$ confidence threshold. The green dotted vertical line shows the true value, and the purple solid line the MLE.}
    \label{fig:profile_likelihood_r0}
\end{figure}

\begin{figure}[!htb]
    \centering
    \begin{subfigure}[t]{0.8\textwidth}
        \centering
        \includegraphics[width=\textwidth]{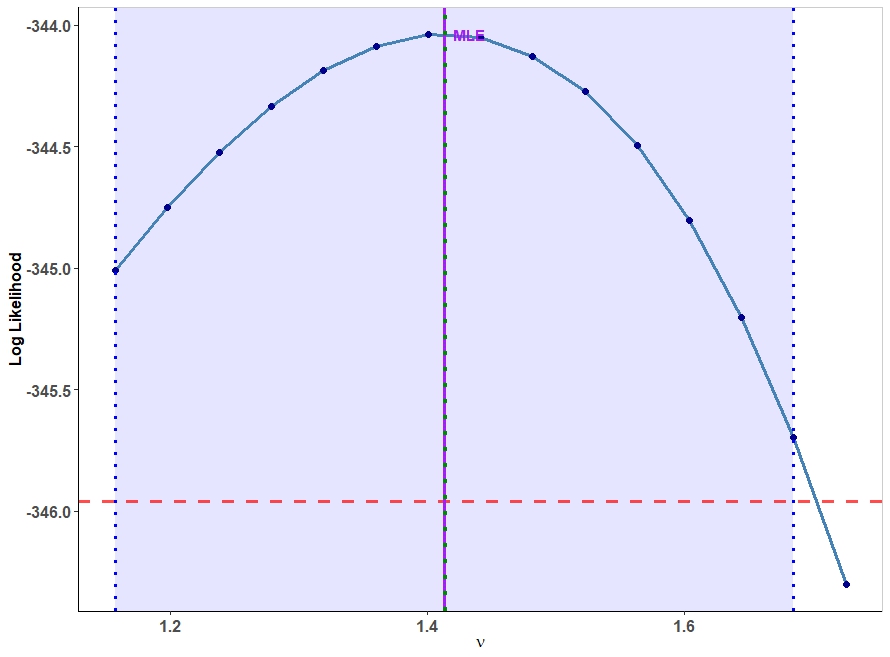}
    \end{subfigure}    
    \begin{subfigure}[t]{0.8\textwidth}
        \centering
        \includegraphics[width=\textwidth]{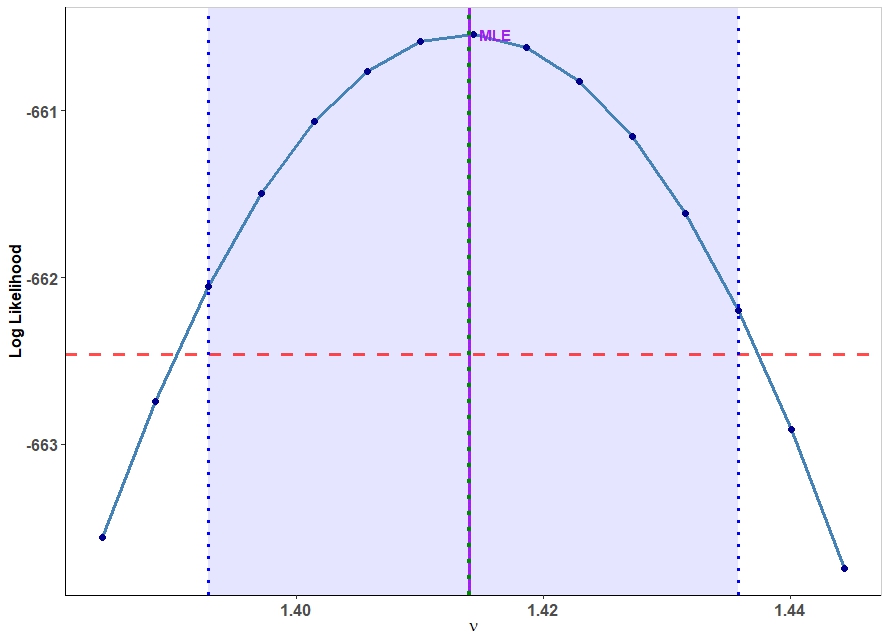}
    \end{subfigure}
    \caption{Profile likelihood curves for the coefficient of variation $\nu$ using single-epidemic (top) and two-epidemic (bottom) models. Horizontal dashed red lines represent the $95\%$ confidence threshold. The green dotted vertical line shows the true value, and the purple solid line the MLE.}
    \label{fig:profile_likelihood_v}
\end{figure}

\begin{figure}[!htb]
    \centering
    \begin{subfigure}[t]{0.8\textwidth}
        \centering
        \includegraphics[width=\textwidth]{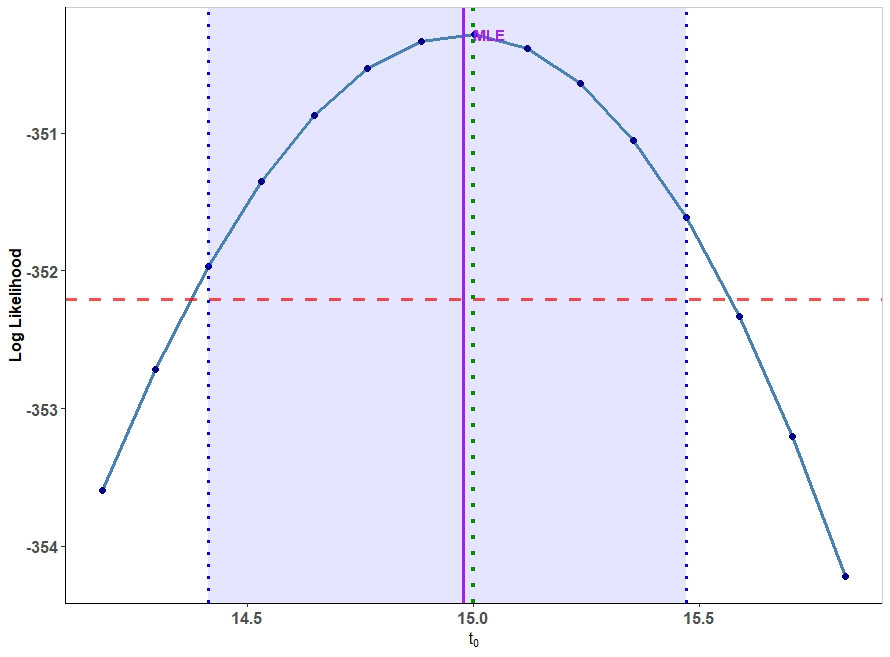}
    \end{subfigure}   
    \begin{subfigure}[t]{0.8\textwidth}
        \centering
        \includegraphics[width=\textwidth]{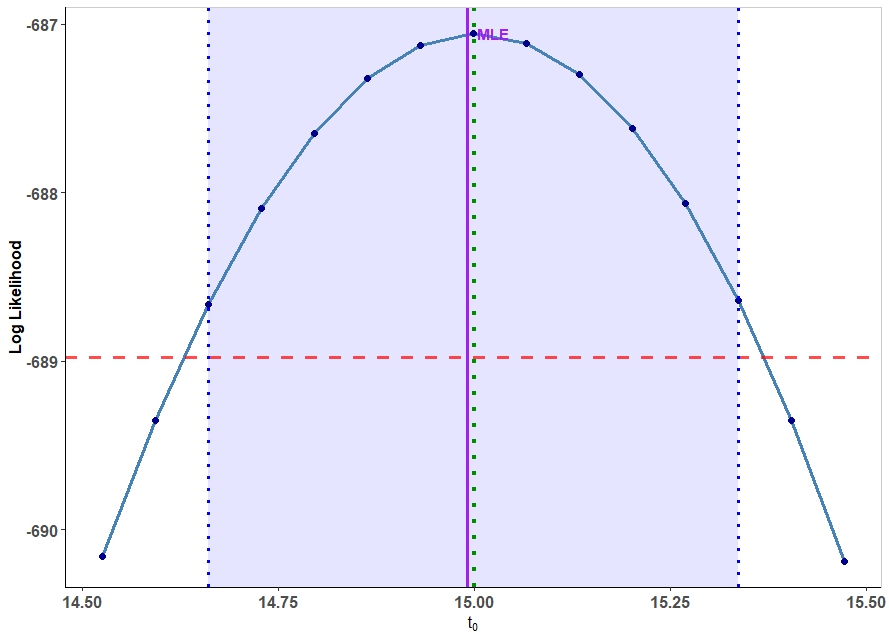}
    \end{subfigure}  
    \caption{Profile likelihood curves for behavioural change timing $t_0$ under single-epidemic (top) and two-epidemic (bottom) models. Horizontal dashed red lines represent the $95\%$ confidence threshold. The green dotted vertical line shows the true value, and the purple solid line the MLE.}
    \label{fig:profile_likelihood_t0}
\end{figure}

\begin{figure}[!htb]
    \centering
    \begin{subfigure}[t]{0.8\textwidth}
        \centering
        \includegraphics[width=\textwidth]{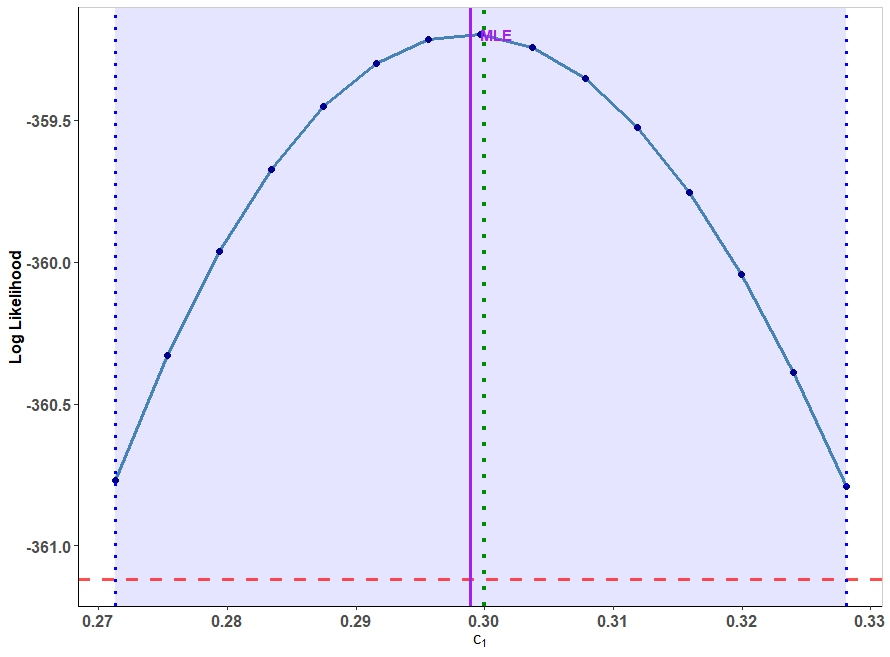}
    \end{subfigure}   
    \begin{subfigure}[t]{0.8\textwidth}
        \centering
        \includegraphics[width=\textwidth]{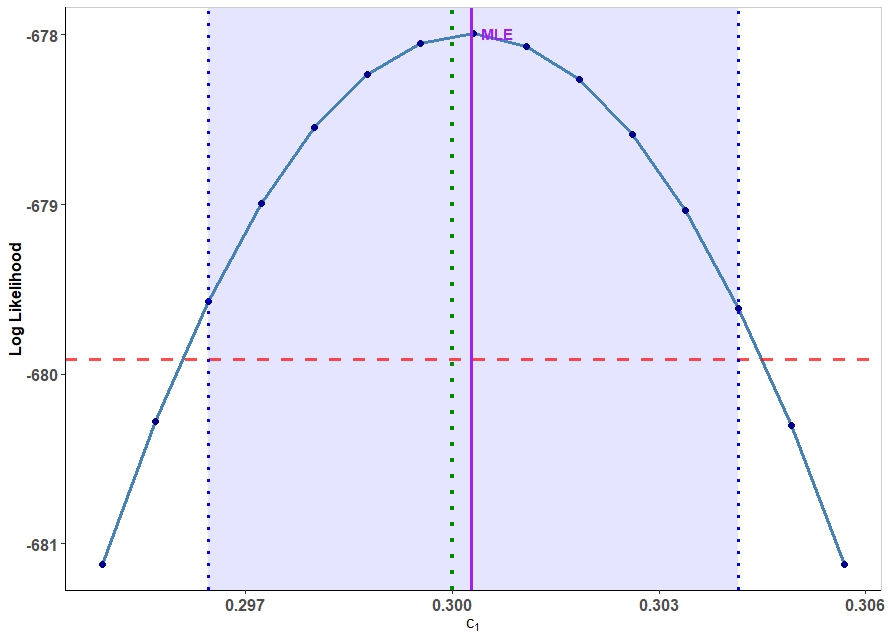}
    \end{subfigure}   
    \caption{Profile likelihood curves for intervention strength $c_1$ under single-epidemic (top) and two-epidemic (bottom) models. Horizontal dashed red lines represent the $95\%$ confidence threshold. The green dotted vertical line shows the true value, and the purple solid line the MLE.}
    \label{fig:profile_likelihood_c}
\end{figure}

Figures~\ref{fig:profile_likelihood_r0}–\ref{fig:profile_likelihood_c} show representative profile likelihood curves for all four key parameters from selected sample dataset, comparing the single-epidemic versus two-epidemic approaches. Profile likelihood curves reveal substantially narrower confidence intervals in the two-epidemic approach compared to the single-epidemic approach across all parameters. This is particularly evident for the coefficient of variation parameter ($\nu$), where the confidence interval width is reduced by approximately $93\%$, and for the intervention strength parameter ($c_1$), where the confidence interval narrows by approximately $86\%$ (Table 3 in the main text).

\FloatBarrier 

\FloatBarrier 

\FloatBarrier 

\FloatBarrier 

\subsection{Accuracy and precision}



\begin{figure}[!htb]
    \centering
    \includegraphics[width=0.9\textwidth]{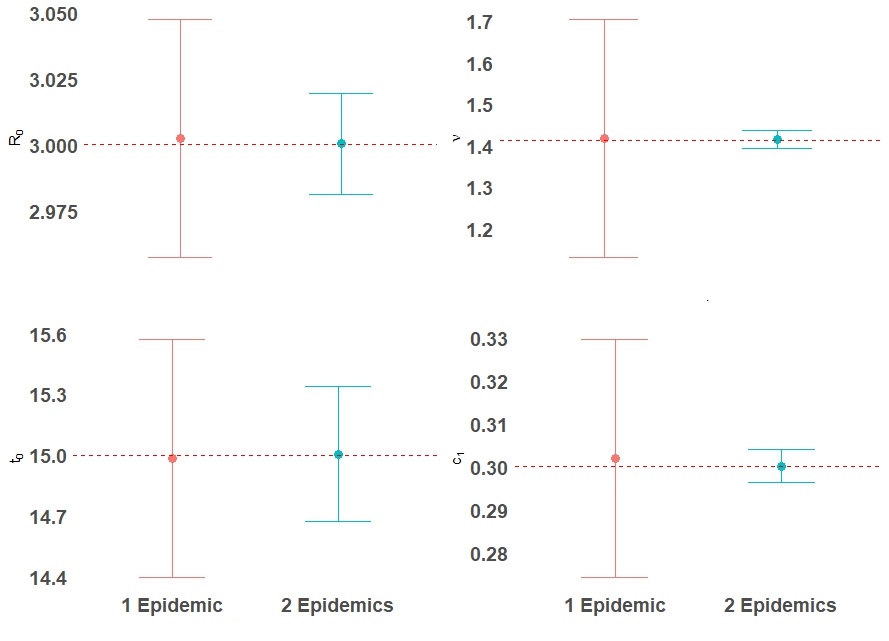}
    \caption{Comparison of parameter estimates between single-epidemic and two-epidemic approaches. Error bars represent the average $95\%$ confidence intervals.}
    \label{fig:parameter_estimates}
\end{figure}

Figure~\ref{fig:parameter_estimates} displays the median and $95\%$ confidence intervals of parameter estimates from all the datasets. The results demonstrate that the two-epidemic approach provides more precise estimates (narrower confidence intervals) and lower relative bias for most parameters. 

\FloatBarrier

Tables~\ref{tab:statistical_performance_1} and \ref{tab:statistical_performance_2} further examine the statistical performance of single and two-epidemic approaches, including relative bias and coverage probabilities of the confidence intervals across all 200 datasets.

\begin{table}[!htb]
\centering
\caption{Statistical performance metrics for the single-epidemic approach.}
\begin{tabular}{lccc}
\toprule
\multirow{2}{*}{Parameter} & \multicolumn{3}{c}{Single epidemic} \\
\cmidrule{2-4}
& Relative bias (\%) & Relative width (\%) & Coverage (\%) \\
\midrule
$\Ro$ & $0.086$ & $3.01$ & $90.36$ \\
$\nu$ & $0.303$ & $40.41$ & $89.34$ \\
$t_0$ & $-0.112$ & $7.83$ & $89.85$ \\
$c_1$ & $0.624$ & $18.46$ & $90.36$ \\
\bottomrule
\end{tabular}
\label{tab:statistical_performance_1}
\end{table}
\FloatBarrier

\begin{table}[!htb]
\centering
\caption{Statistical performance metrics for the two-epidemic approach.}
\begin{tabular}{lccc}
\toprule
\multirow{2}{*}{Parameter} & \multicolumn{3}{c}{Two epidemics} \\
\cmidrule{2-4}
& Relative bias (\%) & Relative width (\%) & Coverage (\%) \\
\midrule
$\Ro$ & $0.014$ & $1.27$ & $93.0$ \\
$\nu$ & $0.038$ & $3.04$ & $93.5$ \\
$t_0$ & $0.036$ & $4.44$ & $93.0$ \\
$c_1$ & $0.052$ & $2.56$ & $91.0$ \\
\bottomrule
\end{tabular}
\label{tab:statistical_performance_2}
\end{table}
\FloatBarrier

\subsection{Numerical stability}

\begin{figure}[!htb]
    \centering
    \includegraphics[width=0.7\textwidth]{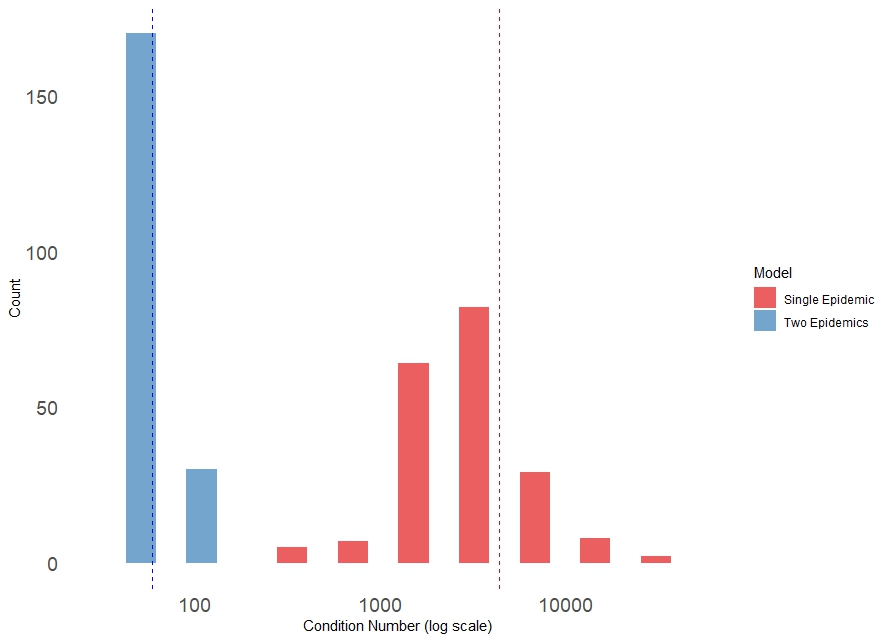}
    \caption{Distribution of Hessian matrix condition numbers (log scale) comparing single and two-epidemic approaches.}
    \label{fig:condition_histogram}
\end{figure}

To assess numerical stability, we computed the condition numbers of the Hessian matrices at the maximum likelihood estimates. Table~\ref{tab:condition_numbers} and Supplementary Figure \ref{fig:condition_histogram} summarize these results for the single and two-epidemic approaches.

\begin{table}[htbp]
\centering
\caption{Hessian Matrix Condition Numbers Comparison}
\begin{tabular}{lcc}
\toprule
Metric & Single epidemic & Two epidemics \\
\midrule
Mean & 4347.5 & 59.2 \\
Median & 3069.6 & 59.3 \\
Minimum & 376.5 & 47.7 \\
Maximum & 40911.3 & 69.3 \\
Standard deviation & 4703.0 & 3.2 \\
\midrule
Relative improvement & \multicolumn{2}{c}{98.6\% reduction} \\
\bottomrule
\end{tabular}
\label{tab:condition_numbers}
\end{table}

\FloatBarrier

The two-epidemic approach exhibits substantially improved numerical stability, with a mean condition number of $59.3$ compared to $3069.6$ for single-epidemic. This indicates that the two-epidemic approach produces a better-conditioned optimization problem. The substantially lower variability in condition numbers (SD of $3.2$ versus $4703.0$) further demonstrates the consistent numerical stability of the two-epidemic approach across different datasets. For scenarios with condition numbers exceeding $40,000$ in the single-epidemic fits, parameters become increasingly difficult to estimate accurately.

\section{Sensitivity analysis}

\begin{figure}[htbp]
    \centering
    \includegraphics[trim={0 0cm 0 1.5cm},clip,width=1\textwidth]{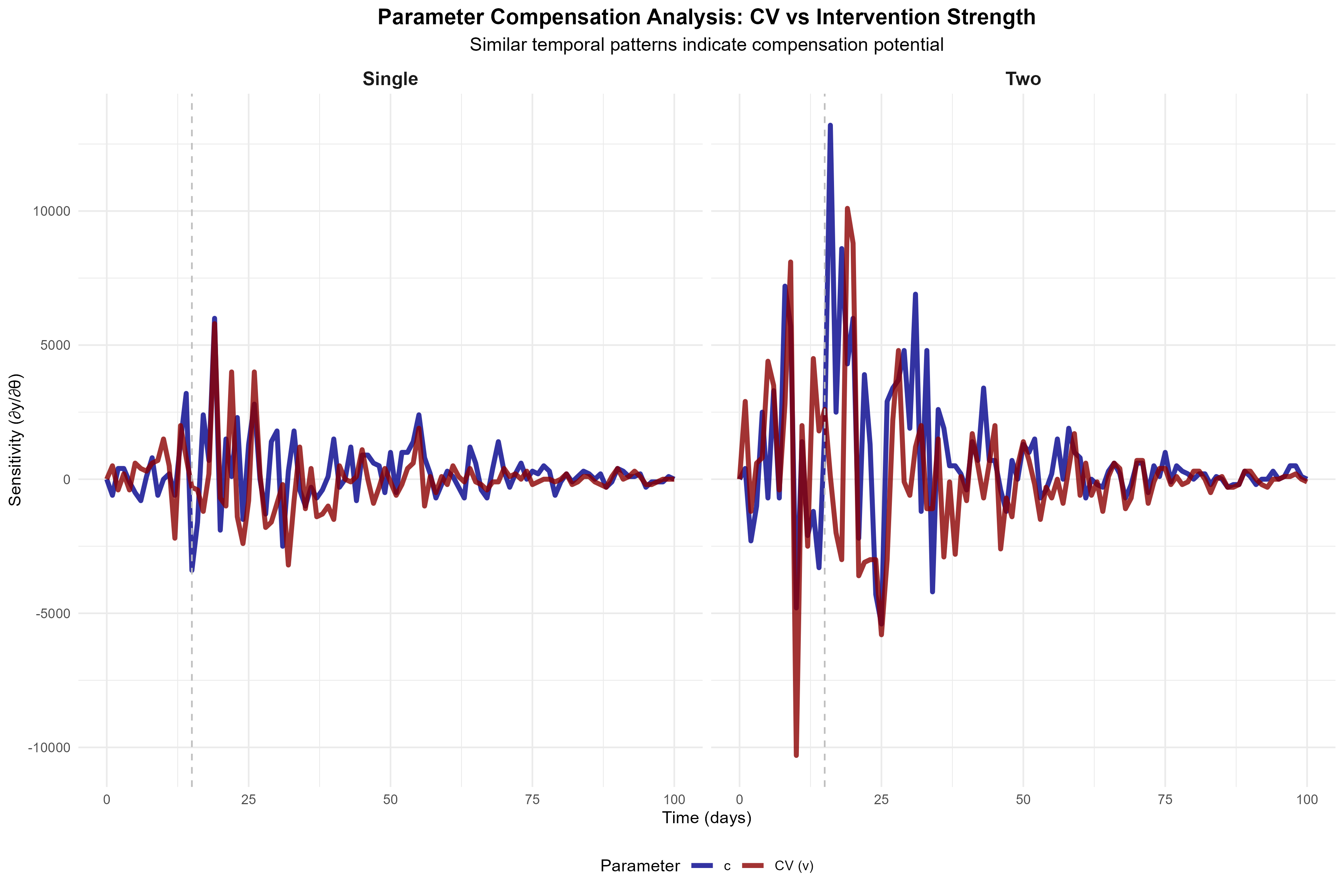}
    \caption{Parameter compensation analysis comparing single versus two-epidemic approaches. Left panel shows highly correlated sensitivity patterns for CV ($\nu$) and intervention strength ($c_1$) in single epidemics, while right panel demonstrates orthogonal patterns in two epidemics that break parameter compensation effects.}
\label{fig:sensitivity_compensation_analysis}
\end{figure}